\pdfoutput=1

\documentclass[aps,prd,amsmath,superscriptaddress,%
               nobibnotes,nofootinbib,preprintnumbers,
               onecolumn,12pt,tightenlines]{revtex4}

\usepackage{graphicx}

\usepackage{ifthen}

\graphicspath{{figures/}}

\newcommand{\ie}{\textit{i.e.}}
\newcommand{\eg}{\textit{e.g.}}
\newcommand{\etal}{\textit{et~al.}}
\newcommand{\nb}{\textit{n.b.}}

\newcommand{\refeqn}[2][eqn:]{Eqn.~(\ref{#1#2})}
\newcommand{\Refeqn}[2][eqn:]{Equation~(\ref{#1#2})}

\newcommand{\reffig}[2][fig:]{Figure~\ref{#1#2}}
\newcommand{\Reffig}[2][fig:]{Figure~\ref{#1#2}}
\newcommand{\refsec}[2][sec:]{Section~\ref{#1#2}} 
\newcommand{\refapp}[2][sec:]{Appendix~\ref{#1#2}}
\newcommand{\Refapp}[2][sec:]{Appendix~\ref{#1#2}}

\newcommand{\ifmulticol}[2]{%
  \ifthenelse{\lengthtest{1.9\columnwidth<\textwidth}}{#1}{#2}%
}

\newcommand{\insertfig}[2][\scfigwidth]{%
    \hspace*{\stretch{1}}
    \includegraphics[keepaspectratio,width=#1\columnwidth]{#2}
    \hspace*{\stretch{1}}
}

\newcommand{\insertwidefig}[2][\widefigwidth]{%
    \hspace*{\stretch{1}}
    \includegraphics[keepaspectratio,width=#1\textwidth]{#2}
    \hspace*{\stretch{1}}
}

\newcommand{\rmd}{\ensuremath{\mathrm{d}}}

\newcommand{\orderof}[1]{\ensuremath{\mathcal{O}(#1)}}
\newcommand{\erf}{\mathop{\mathrm{erf}}}

\newcommand{\gae}{%
  \ensuremath{\raisebox{-0.4ex}{%
    $\,\buildrel{\scriptstyle>}\over{\scriptstyle\sim}\,$}%
    }%
  }
\newcommand{\lae}{%
  \ensuremath{\raisebox{-0.4ex}{%
    $\,\buildrel{\scriptstyle<}\over{\scriptstyle\sim}\,$}%
    }%
  }

\newcommand{\Enr}{\ensuremath{E_{\mathrm{nr}}}}
\newcommand{\Eee}{\ensuremath{E_{\mathrm{ee}}}}

\newcommand{\dRdE}{\ensuremath{\frac{dR}{dE}}}

\newcommand{\dRdEnr}{\ensuremath{\frac{dR}{dE_{\mathrm{nr}\!\!\!\!\!}}\;}}
\newcommand{\dRidEnr}{\ensuremath{\frac{dR_{i\!\!}}{dE_{\mathrm{nr}\!\!\!\!\!}}\;}}
\newcommand{\dRMDidEnr}{\ensuremath{\frac{dR_{i\!\!}^{\mathrm{MD}\!\!\!\!\!\!\!\!\!\!}}{dE_{\mathrm{nr}\!\!\!\!\!}}\;\;\;}}
\newcommand{\dRdEee}{\ensuremath{\frac{dR}{dE_{\mathrm{ee}\!\!\!\!\!}}\;}}

\newcommand{\mchi}{\ensuremath{m_{\chi}}}
\newcommand{\rhochi}{\ensuremath{\rho_{\chi}}}
\newcommand{\nchi}{\ensuremath{n_{\chi}}}

\newcommand{\vmin}{\ensuremath{v_\mathrm{min}}}
\newcommand{\vmp}{\ensuremath{v_0}}
\newcommand{\vrot}{\ensuremath{v_\mathrm{rot}}}
\newcommand{\vobs}{\ensuremath{v_\mathrm{obs}}}
\newcommand{\bvobs}{\ensuremath{\mathbf{v}_\mathrm{obs}}}
\newcommand{\vesc}{\ensuremath{v_\mathrm{esc}}}
\newcommand{\Nesc}{\ensuremath{N_\mathrm{esc}}}
\newcommand{\vflow}{\ensuremath{v_\mathrm{flow}}}
\newcommand{\bv}{\ensuremath{\mathbf{v}}}  
\newcommand{\bV}{\ensuremath{\mathbf{V}}}  
\newcommand{\eone}{\ensuremath{\hat{\boldsymbol{\varepsilon}}_1}}  
\newcommand{\etwo}{\ensuremath{\hat{\boldsymbol{\varepsilon}}_2}}  
\newcommand{\egen}{\ensuremath{\hat{\boldsymbol{\varepsilon}}_i}}  

\newcommand{\qmax}{\ensuremath{q_{\mathrm{max}}}}  

\newcommand{\fpSI}{\ensuremath{f_{\mathrm{p}}}}
\newcommand{\fnSI}{\ensuremath{f_{\mathrm{n}}}}
\newcommand{\apSD}{\ensuremath{a_{\mathrm{p}}}}
\newcommand{\anSD}{\ensuremath{a_{\mathrm{n}}}}

\newcommand{\sigmaSI}{\ensuremath{\sigma_{\mathrm{SI}}}}
\newcommand{\sigmaSD}{\ensuremath{\sigma_{\mathrm{SD}}}}

\newcommand{\sigmapSI}{\ensuremath{\sigma_{\mathrm{p,SI}}}}
\newcommand{\sigmanSI}{\ensuremath{\sigma_{\mathrm{n,SI}}}}

\newcommand{\sigmapSD}{\ensuremath{\sigma_{\mathrm{p,SD}}}}
\newcommand{\sigmanSD}{\ensuremath{\sigma_{\mathrm{n,SD}}}}

\newcommand{\mup}{\ensuremath{\mu_{\mathrm{p}}}}

\newcommand{\Sp}{\ensuremath{\langle S_{\mathrm{p}} \rangle}}
\newcommand{\Sn}{\ensuremath{\langle S_{\mathrm{n}} \rangle}}

\newcommand{\Scp}{\ensuremath{S_0^{\,^\prime}}}
\newcommand{\Smp}{\ensuremath{S_m^{\,^\prime}}}
\newcommand{\Smpsq}{\ensuremath{S_m^{\,^\prime\,2}}}

\begin{document}


\preprint{MCTP-12-15}


\title{Annual Modulation of Dark Matter: A Review}

\author{Katherine Freese}
\email[]{ktfreese@umich.edu}
\affiliation{
 Michigan Center for Theoretical Physics,
 Department of Physics,
 University of Michigan,
 Ann Arbor, MI 48109}
\affiliation{
Physics Department,
Caltech,
Pasadena, CA 91101}

\author{Mariangela Lisanti}
\email[]{mlisanti@princeton.edu}
\affiliation{
 Princeton Center for Theoretical Science,
 Princeton University,
 Princeton, NJ 08544}

\author{Christopher Savage}
\email[]{savage@physics.utah.edu}
\affiliation{
 The Oskar Klein Centre for Cosmoparticle Physics,
 Department of Physics,
 Stockholm University,
 AlbaNova,
 SE-106 91 Stockholm, Sweden}
\affiliation{
 Department of Physics \& Astronomy,
 University of Utah,
 Salt Lake City, UT 84112}

\date{\today}



\begin{abstract} 

\vspace{0.2in}
\begin{center}
  \textbf{Abstract}
\end{center}
Direct detection experiments, which are designed to detect the
scattering of dark matter off nuclei in detectors, are a critical
component in the search for the Universe's missing matter.  The count
rate in these experiments should experience an annual modulation due
to the relative motion of the Earth around the Sun. This modulation,
not present for most known background sources, is critical for
solidifying the origin of a potential signal as dark matter.  In this
article, we review the physics of annual modulation, discussing the
practical formulae needed to interpret a modulating signal.  We focus
on how the modulation spectrum changes depending on the particle and
astrophysics models for the dark matter.  For standard assumptions,
the count rate has a cosine dependence with time, with a maximum in
June and a minimum in December.  Well-motivated generalizations of
these models, however, can affect both the phase and amplitude of the
modulation.  We show how a measurement of an annually modulating
signal could teach us about the presence of substructure in the
Galactic halo or about the interactions between dark and baryonic
matter.  Although primarily a theoretical review, we briefly discuss the
current experimental situation for annual modulation and future
experimental directions.

\end{abstract} 

\maketitle


\pagebreak

\section{\label{sec:Intro} Introduction}

The Milky Way galaxy is known to be surrounded by a halo of dark
matter whose composition remains a mystery.  Only 5\% of the Universe
consists of ordinary atomic matter, while the remainder is 23\% dark
matter and 72\% dark energy \cite{Komatsu:2010fb}.  Identifying the
nature of this dark matter is the longest outstanding problem in all
of modern physics, stemming back to observations in 1933 by Fritz
Zwicky; he proposed the existence of ``Dunkle Materie'' (German for
``dark matter'') as a source of gravitational potential to explain
rapid motions of galaxies in the Coma Cluster~\cite{Zwicky:1937zza}.
Subsequently, others discovered flat rotation curves in disk galaxies,
starting with Babcock in 1939~\cite{Babcock:1939} and followed (more
persuasively and with better data) by Rubin and Ford
~\cite{Rubin:1970zza} and Roberts and Whitehurst~\cite{Roberts:1975}
in the 1970s.  Their results imply that the predominant constituent of
mass inside galaxies must be nonluminous matter (see
Refs.~\cite{Sandage:1975,Faber:1979pp} for reviews).

A leading candidate for this dark matter is a Weakly Interacting
Massive Particle (WIMP).  The terminology refers to the fact that
these particles undergo weak interactions in addition to feeling the
effects of gravity, but do not participate in electromagnetic or
strong interactions.  WIMPs are electrically neutral and the average
number of interactions with the human body is at most one per minute,
even with billions passing through every second~\cite{Freese:2012rp}.
The expected WIMP mass ranges from 1~GeV to 10~TeV.  These particles,
if present in thermal equilibrium in the early universe, annihilate
with one another so that a predictable number of them remain today.
The relic density of these particles is
\begin{equation}
  \Omega_\chi h^2 \sim (3 \times 10^{-26} \mathrm{cm}^3/\mathrm{sec})
                        / \langle \sigma v \rangle_{\mathrm{ann}},
\end{equation}
where $\Omega_{\chi}$ is the fractional contribution of WIMPs to the
energy density of the Universe.  An annihilation cross section
$\langle \sigma v \rangle_{\mathrm{ann}}$ of weak interaction strength
automatically gives the right answer, near the value measured by WMAP
\cite{Komatsu:2010fb}.  This coincidence is known as the ``WIMP
miracle'' and is why WIMPs are taken so seriously as dark matter
candidates.  Possibly the best WIMP candidate is motivated by
supersymmetry (SUSY): the lightest neutralino in the Minimal
Supersymmetric Standard Model (MSSM) and its
extensions~\cite{Jungman:1995df}.  However, other WIMP candidates
arise in a variety of theories beyond the Standard Model (see
Refs.~\cite{Bergstrom:2000pn,Bertone:2004pz} for a review).

A multitude of experimental efforts are currently underway to detect
WIMPs, with some claiming hints of detection.  There is a
three-pronged approach: particle accelerator, indirect detection
(astrophysical), and direct detection experiments.  The focus of this
review article is the third option -- direct detection experiments.
This field began thirty years ago with the work of Drukier and
Stodolsky \cite{Drukier:1983gj}, who proposed searching for weakly
interacting particles (with a focus on neutrinos) by observing the
nuclear recoil caused by their weak interactions with nuclei in
detectors.  Then, Goodman and Witten \cite{Goodman:1984dc} made the
important point that this approach could be used to search not just
for neutrinos but also for WIMPs, again via their weak interactions
with detectors.  Soon after, Drukier, Freese, and Spergel
\cite{Drukier:1986tm} extended this work by taking into account the
halo distribution of WIMPs in the Milky Way, as well as proposing the
annual modulation that is the subject of this review.

The basic goal of direct detection experiments is to measure the
energy deposited when WIMPs interact with nuclei in a detector,
causing those nuclei to recoil.  The experiments, which are typically
located far underground to reduce background contamination, are
sensitive to WIMPs that stream through the Earth and interact with
nuclei in the detector target.  The recoiling nucleus can deposit
energy in the form of ionization, heat, and/or light that is
subsequently detected.  In the mid 1980s, the development of
ultra-pure germanium detectors provided the first limits on WIMPs
\cite{Ahlen:1987mn}.  Since then, numerous collaborations worldwide
have been (or will be) searching for these particles, including
ANAIS \cite{Amare:2011zz},
ArDM \cite{Marchionni:2010fi},
CDEX/TEXONO \cite{Wong:2010zz},
CDMS \cite{Akerib:2005zy,Ahmed:2009zw,Ahmed:2010wy,Ahmed:2012vq},
CoGeNT \cite{Aalseth:2012if,Aalseth:2010vx,Aalseth:2011wp},
COUPP \cite{Behnke:2012ys},
CRESST \cite{Angloher:2011uu},
DAMA/NaI \cite{Bernabei:2003za},
DAMA/LIBRA \cite{Bernabei:2008yh,Bernabei:2010mq},
DEAP/CLEAN \cite{Kos:2010zz},
DM-Ice \cite{Cherwinka:2011ij},
DRIFT \cite{Alner:2005xp,Daw:2010ud},
EDELWEISS \cite{Sanglard:2005we,Armengaud:2011cy,Armengaud:2012pfa},
EURECA \cite{Kraus:2011zz},
KIMS \cite{Kim:2012rz},
LUX \cite{Hall:2010zz},
NAIAD \cite{Alner:2005kt},
PandaX \cite{Li:2012zs},
PICASSO \cite{BarnabeHeider:2005ri,Archambault:2012pm},
ROSEBUD \cite{Coron:2011zz},
SIMPLE \cite{Felizardo:2011uw},
TEXONO \cite{Lin:2007ka},
WArP \cite{Acciarri:2011zz},
XENON10 \cite{Aprile:2010bt,Angle:2007uj,Angle:2011th},
XENON100 \cite{Aprile:2011dd,Aprile:2012nq},
XENON1T \cite{Aprile:2012zx},
XMASS \cite{Moriyama:2011zz},
ZEPLIN \cite{Akimov:2006qw,Akimov:2011tj},
and many others.

\begin{figure}
	\insertfig{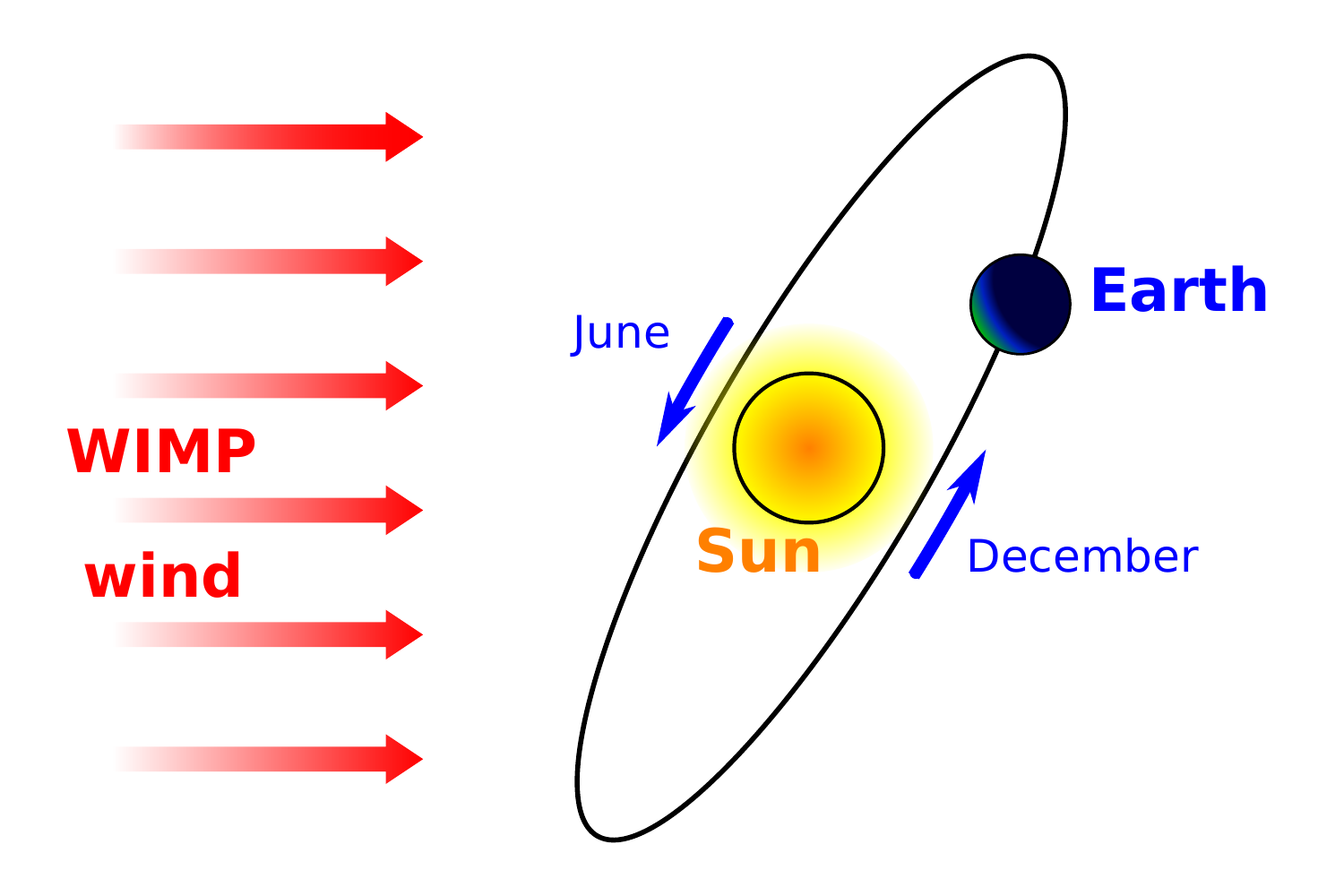}
    \caption{
      A simplified view of the WIMP velocities as seen from the Sun
      and Earth.  Due to the rotation of the Galactic Disk (containing
      the Sun) through the essentially non-rotating dark matter halo,
      the solar system experiences an effective ``WIMP wind.''  From
      the perspective of the Earth, the wind changes throughout the
      year due to the Earth's orbital motion: the wind is at maximum
      speed around the beginning of June, when the Earth is moving
      fastest in the direction of the disk rotation, and at a minimum
      speed around the beginning of December, when the Earth is moving
      fastest in the direction opposite to the disk rotation.  The
      Earth's orbit is inclined at $\sim$60$^{\circ}$ relative to the
      plane of the Disk.
      }
  \label{fig:earth}
\end{figure}

The count rate in direct detection experiments experiences an annual
modulation \cite{Drukier:1986tm,Freese:1987wu} due to the motion of
the Earth around the Sun (see \reffig{earth}).  Because the relative
velocity of the detector with respect to the WIMPs depends on the time
of year, the count rate exhibits (in most cases) a sinusoidal
dependence with time.  For the simplest assumptions about the dark
matter distribution in the halo, the flux is maximal in June and
minimal in December.  Annual modulation is a powerful signature for
dark matter because most background signals, \eg\ from radioactivity
in the surroundings, are not expected to exhibit this kind of time
dependence.  The details concerning the recoil energy and modulation
spectra depend on the specifics of both the particle physics model and
the distribution of WIMPs in the Galaxy.  We discuss these
possibilities in this review.

For more than a decade, the DAMA experiment \cite{Bernabei:2008yh} has
been claiming detection of an annual modulation.  The experiment,
which consists of NaI crystals, is situated in the Gran Sasso Tunnel
under the Apennine Mountains near Rome.  By now, the amount of data
collected is enormous and the statistical significance of the result
is undeniable.  The DAMA annual modulation is currently reported as
almost a 9$\sigma$ effect \cite{Bernabei:2010mq}, and is consistent
with an $\sim$80 or 10~GeV~\cite{Bottino:2003iu,
  Bottino:2003cz,Gondolo:2005hh,Petriello:2008jj,Chang:2008xa,
  Savage:2008er} WIMP elastically scattering predominantly off of
iodine or sodium, respectively.  Many other direct detection
experiments have presented null results that are in clear conflict
with the high-mass window.  The viability of the 10~GeV WIMP remains a
controversial issue because it is not clearly compatible nor clearly
incompatible with other null experiments once various detector
systematics are taken into account.  Recently, the CoGeNT experiment
reported a 2.8$\sigma$ evidence for an annual
modulation~\cite{Aalseth:2011wp} and a third experiment, CRESST-II,
has also announced anomalous results~\cite{Angloher:2011uu}.  Whether
DAMA, CoGeNT, and CRESST are consistent in the low-mass window is
still debated~\cite{Kelso:2011gd,Fox:2011px}.  Yet CDMS sees no annual
modulation~\cite{Ahmed:2012vq}, and both CDMS
\cite{Ahmed:2009zw,Ahmed:2010wy} and
XENON~\cite{Angle:2011th,Aprile:2012nq} find null results that appear
to be in conflict with the three experiments that report anomalies.

The current experimental situation in direct detection searches is
exciting.  Understanding the anomalies and the role that different
experiments play in validating them is of crucial importance in moving
forward in the search for dark matter.  In this review article, we
seek to provide the reader with the basic theoretical tools necessary to
understand a potential dark matter signature at a direct detection
experiment, focusing on the annual modulation of the signal.  We begin
in \refsec{DMDetection} by reviewing the basics of direct detection
techniques for WIMPs, describing the particle physics in
\refsec{CrossSection} and the astrophysics in \refsec{VelocityDist}.
We describe the Standard Halo Model (SHM) as well as modifications due
to substructures. In \refsec{Modulation}, we examine the behavior of
the annual modulation signals for both the SHM and substructures.
Although this is primarily a theoretical review, we turn to the
experimental status in \refsec{Experiments}, briefly reviewing the
current anomalies and null results.  We conclude in \refsec{Summary}.
The Appendices discuss quantities required for understanding results of
direct detection experiments.
\Refapp{Quench} describes the quenching factor, and \refapp{MIV}
presents analytical results for the mean inverse speed for commonly
used WIMP velocity distributions, a quantity necessary for a
computation of expected count rates in detectors.

\section{\label{sec:DMDetection} Dark Matter Detection}

Direct detection experiments aim to observe the recoil of a nucleus in
a collision with a dark matter particle~\cite{Goodman:1984dc}.  After
an elastic collision with a WIMP $\chi$ of mass $\mchi$, a nucleus of
mass $M$ recoils with energy $\Enr = (\mu^2 v^2/M)(1-\cos\theta)$,
where $\mu \equiv \mchi M/ (\mchi + M)$ is the reduced mass of the
WIMP-nucleus system, $v$ is the speed of the WIMP relative to the
nucleus, and $\theta$ is the scattering angle in the center of mass
frame.  The differential recoil rate per unit detector mass is
\begin{equation}\label{eqn:dRdEnr}
  \dRdEnr
    = \frac{\nchi}{M} \Big\langle v \frac{d\sigma}{d\Enr} \; \Big\rangle
    = \frac{2\rhochi}{\mchi}
      \int d^3v \, v f(\bv,t) \frac{d\sigma}{dq^2}(q^2,v) \, ,
\end{equation}
where $\nchi = \rhochi/\mchi$ is the number density of WIMPs, with
$\rhochi$ the local dark matter mass density; $f(\bv,t)$ is the
time-dependent WIMP velocity distribution; and
$\frac{d\sigma}{dq^2}(q^2,v)$ is the velocity-dependent differential
cross-section, with $q^2 = 2 M \Enr$ the momentum exchange in the
scatter.  The differential rate is typically given in units of
cpd\,kg$^{-1}$\,keV$^{-1}$, where cpd is counts per day.  Using the
form of the differential cross-section for the most commonly assumed
couplings, to be discussed below,
\begin{equation}\label{eqn:dRdEnr2}
  \dRdEnr
    = \frac{1}{2 \mchi \mu^2} \, \sigma(q) \, \rhochi \eta(\vmin(\Enr),t),
\end{equation}
where $\sigma(q)$ is an effective scattering cross-section and
\begin{equation} \label{eqn:eta}  
  \eta(\vmin,t) = \int_{v > \vmin} d^3v \, \frac{f(\bv,t)}{v}
\end{equation}
is the mean inverse speed, with
\begin{equation} \label{eqn:vmin}
  \vmin =
    \begin{cases}
      \sqrt{\frac{M \Enr}{2\mu^2}}
        & \textrm{(elastic)} \\
      \frac{1}{\sqrt{2 M \Enr}} \Big(\frac{M \Enr}{\mu} + \delta\Big)
      \qquad
        & \textrm{(inelastic)}
    \end{cases}
\end{equation}
the minimum WIMP velocity that can result in a recoil energy $\Enr$.
Here, $\delta$ is the mass splitting between the lightest and
next-to-lightest states in the spectrum in the case of an inelastic
scattering interaction;\footnote{%
  Inelastic scattering with $\delta \simeq \mathcal{O}$(100~keV) was
  first invoked to reconcile the DAMA anomaly with the CDMS
  limits~\cite{TuckerSmith:2001hy}.  Although this explanation has
  since been ruled out by XENON100 for conventional couplings
  \cite{Aprile:2011ts}, alternate formulations remain viable as a
  means of reconciling the experimental results (see
  \eg\ Ref.~\cite{Chang:2010en}).  More generally, inelastic
  scattering (for arbitrary $\delta$) remains an interesting
  possibility for direct detection experiments, yielding distinct
  recoil spectra.
  }
we shall consider only the elastic scattering case for the remainder
of this review.  The benefit of writing the recoil spectrum in the
form of \refeqn{dRdEnr2} is that the particle physics and astrophysics
separate into two factors, $\sigma(q)$ and $\rhochi \eta(\vmin,t)$,
respectively.\footnote{%
  The ability to separate the particle physics and astrophysics terms
  in the manner shown requires that the differential scattering
  cross-section $\frac{d\sigma}{dq^2}$ be of the form presented 
  in \refeqn{dsigmadq}.
  While this form is expected for the most common types of interactions
  studied, there are other interactions for which the particle physics
  and astrophysics cannot be separated as described.
  }
We discuss each of these factors in the following two
subsections.  More detailed reviews of the dark matter scattering
process and direct detection can be found in
Refs.~\cite{Primack:1988zm,Smith:1988kw,Lewin:1995rx,Jungman:1995df,
Bertone:2004pz}.

\subsection{\label{sec:CrossSection} Particle Physics: Cross-section}

For a SUSY neutralino and many other WIMP candidates, the dominant
WIMP-quark couplings in direct detection experiments are the scalar
and axial-vector couplings, which respectively give rise to
spin-independent (SI) and spin-dependent (SD)
cross-sections~\cite{Jungman:1995df}.  In both cases,
\begin{equation}\label{eqn:dsigmadq}
  \frac{d\sigma}{dq^2}(q^2,v) = \frac{\sigma_{0}}{4 \mu^2 v^2}
                              F^2(q) \, \Theta(\qmax-q)
\end{equation}
to leading order (see \eg\ \cite{Cirigliano:2012pq} for how higher order
corrections can modify this form).
Here, $\Theta$ is the Heaviside step function, $\qmax = 2 \mu v$ is
the maximum momentum transfer in a collision at a relative velocity
$v$, and the requirement $q < \qmax$ gives rise to the lower limit $v
> \vmin$ in the integral for $\eta$ in \refeqn{eta}. In the above
equation, $\sigma_0$ is the scattering cross-section in the
zero-momentum-transfer limit---we shall use $\sigmaSI$ and $\sigmaSD$
to represent this term in the SI and SD cases, respectively---and
$F^2(q)$ is a form factor to account for the finite size of the
nucleus.  The WIMP coherently scatters off the entire nucleus when the
momentum transfer is small, giving $F^2(q) \to 1$.  However, as the
de~Broglie wavelength of the momentum transfer becomes comparable to
the size of the nucleus, the WIMP becomes sensitive to the spatial
structure of the nucleus and $F^2(q) < 1$, with $F^2(q) \ll 1$ at
higher momentum transfers.  It is traditional to define a form-factor
corrected cross-section
\begin{equation}\label{eqn:sigmaq}
  \sigma(q) \equiv \sigma_0 F^2(q) \, ,
\end{equation}
as was used in \refeqn{dRdEnr2} above.  We note that this is an
\textit{effective} cross-section, whereas the \textit{actual}
scattering cross-section is given by $\int dq^2
\frac{d\sigma}{dq^2}(q^2,v)$ for a given relative velocity $v$.

The total WIMP-nucleus scattering rate is then the sum over both the
SI and SD contributions, each with its own value of the form factor.
We describe these two cross-sections below and then briefly discuss
more general operators.

\subsubsection{\label{sec:CSSI} Spin-independent cross-section (SI)}

The SI WIMP-nucleus interaction, which occurs through operators such
as $(\bar{\chi}\chi)(\bar{q}q)$, has the cross-section 
\begin{equation} \label{eqn:sigmaSI}
  \sigmaSI = \frac{4}{\pi} \mu^2
             \Big[ Z \fpSI + (A-Z) \fnSI \Big]^{2} \; ,
\end{equation}
where $Z$ and $A-Z$ are the number of protons and neutrons in the
nucleus, respectively, and $\fpSI$ ($\fnSI$) is the effective coupling
to the proton (neutron).  For neutralinos and most other WIMP
candidates with a SI interaction arising through scalar couplings,
$\fnSI \simeq \fpSI$ and the SI scattering cross-section of WIMPs with
protons and neutrons are roughly comparable, $\sigmanSI \approx
\sigmapSI$.  For identical couplings ($\fnSI = \fpSI$), the SI
cross-section can be written as
\begin{equation} \label{eqn:sigmaSI2}
  \sigmaSI = \frac{\mu^2}{\mup^2} A^2 \, \sigmapSI \, ,
\end{equation}
where $\mup$ is the WIMP-proton reduced mass.  As neutralinos are the
currently favored WIMP candidate, this assumption is widely made
throughout the direct detection literature.  Although typically $\fnSI
\simeq \fpSI$, models can be constructed that violate this condition
(\eg, isospin-violating dark matter \cite{Feng:2011vu}).  We assume
identical SI couplings for the rest of this review.

The SI cross-section grows rapidly with nuclear mass.  The explicit
$A^2$ factor in \refeqn{sigmaSI2} arises from the fact that the
contributions to the total SI cross-section of a nucleus is a coherent
sum over the individual protons and neutrons within.  In addition, for
WIMPs that are much heavier than the nucleus, $\frac{\mu^2}{\mup^2}
\approx A^2$, so the cross-section scales as $\sim A^4$.  However, the
form factor suppression becomes more significant as the size of the
nucleus increases, so the scattering rate does not scale as $\sim A^4$
for heavy nuclei, though it still rises rapidly with $A$.  As a
result, direct detection experiments often use heavy nuclei to
increase their sensitivity to WIMP scattering.

The SI form factor is essentially a Fourier transform of the mass
distribution of the nucleus.  A reasonably accurate approximation is
the Helm form factor \cite{Helm:1956zz,Lewin:1995rx}:
\begin{equation} \label{eqn:SIFF}
  F(q) = 3 e^{-q^2 s^2/2} \, \frac{\sin(qr_n)- qr_n\cos(qr_n)}{(qr_n)^3} \, ,
\end{equation}
where $s\simeq 0.9$~fm and $r_n^2 = c^2 + \frac{7}{3} \pi^2 a^2 - 5
s^2$ is an effective nuclear radius with $a \simeq 0.52$~fm and $c
\simeq 1.23 A^{1/3} - 0.60$~fm.  Further details on SI form factors
can be found in Refs.~\cite{Lewin:1995rx,Duda:2006uk}.

\subsubsection{\label{sec:CSSD} Spin-dependent cross-section (SD)}

SD scattering is due to the interaction of a WIMP with the spin of the
nucleus through operators such as $(\bar{\chi}\gamma_{\mu}\gamma_5
\chi)(\bar{q} \gamma^{\mu} \gamma_5 q)$, and takes place only in those
detector isotopes with an unpaired proton and/or unpaired neutron.
The SD WIMP-nucleus cross-section is
\begin{equation} \label{eqn:sigmaSD}
  \sigmaSD = \frac{32 \mu^2}{\pi} G_{F}^{2} J(J+1) \Lambda^2 \, ,
\end{equation}
where $G_F$ is the Fermi constant, $J$ is the spin of the nucleus,
\begin{equation} \label{eqn:Lambda}
  \Lambda \equiv \frac{1}{J} \Big( \apSD \Sp + \anSD \Sn \Big) \, ,
\end{equation}
where $\Sp$ and $\Sn$ are the average spin contributions from the
proton and neutron groups, respectively, and $\apSD$ ($\anSD$) are the
effective couplings to the proton (neutron).  Unlike the SI case, the
two SD couplings $\anSD$ and $\apSD$ may differ substantially (though
they are often of similar order of magnitude), so that a
simplification comparable to \refeqn{sigmaSI2} for SI scattering is
not made in the SD case.  Because of the uncertain theoretical
relation between the two couplings and following from the fact that
one of $\Sp$ or $\Sn$ is often much smaller than the other,
experiments typically only significantly constrain one of the two SD
cross-sections, $\sigmapSD$ or $\sigmanSD$, but not both.

SD scattering is often of lesser significance than SI scattering in
direct detection experiments for two main reasons. First, SI
scattering has a coherence factor $A^2$ that the SD scattering is
missing.  In fact, the spin factors $J$, $\Sp$, and $\Sn$ are either
zero or $\orderof{1}$, so the SD cross-section does not grow as
rapidly with nucleus size as the SI cross-section does.  Thus, whereas
$\sigmaSI \propto A^4$ for heavy WIMPs, $\sigmaSD \propto A^2$ (\nb\
this remaining $A^2$ factor arises from $\frac{\mu^2}{\mup^2} \approx
A^2$).  Second, spin-zero isotopes do not contribute to SD scattering,
so the SD scattering is reduced in elements where non-zero-spin nuclei
represent only a small fraction of the naturally occurring isotopes
within a detector's target mass.  We note that SD couplings may often
be larger than SI couplings; \eg, for an MSSM neutralino, it is often
the case that $\sigmapSD/\sigmapSI \sim \orderof{10^2-10^4}$.
However, even with this ratio of couplings, SI scattering is still
expected to dominate for the heavy elements used in most detectors for
the two reasons described above.

The SD form factor depends on the spin structure of a nucleus and is
thus different between individual elements.  Form factors for many
isotopes of interest to direct detection experiments, as well as
estimates of the spin factors $\Sp$ and $\Sn$, can be found in the
reviews of Refs.~\cite{Bednyakov:2004xq,Bednyakov:2006ux}.

\subsubsection{\label{sec:CSGen} General operators}

While scalar and axial-vector couplings are the dominant interactions
for many WIMP candidates, such as neutralinos, they are by no means
the only allowed couplings.  In general, dark matter-nucleon
interactions can be described by a non-relativistic effective theory,
as detailed in~\cite{Fan:2010gt, Fitzpatrick:2012ix}.  The effective
theory approach is useful for highlighting the variety of operator
interactions that can exist, and their potentially unique direct
detection signatures.

Generic operators can give rise to additional factors of the velocity
and/or momentum in \refeqn{dsigmadq}.  Due to the small velocities
($v \sim 10^{-3}c$) and momenta transfers, these interactions are
expected to be suppressed relative to the scalar and axial-vector cases
and are thus often ignored.  However, in models where the scalar and
axial-vector couplings are forbidden or suppressed themselves, these new
types of interactions can become important.

Consider momentum-dependent (MD) interactions.
For certain classes of theories~\cite{Chang:2009yt,Alves:2009nf,
Masso:2009mu, Feldstein:2009tr, An:2010kc}, the dominant interactions
yield a scattering rate of the form
\begin{equation}
  \dRMDidEnr = \Bigg(\frac{q^2}{q_0^2}\Bigg)^n \dRidEnr \, ,
\end{equation}
where $q_0$ is an arbitrary mass scale and $i =$ SI, SD denotes
whether the rate is independent of nuclear spin or not; $\dRidEnr$ is
the conventional SI or SD scattering rate described previously.  For
the most commonly studied operators, $(\bar{\chi}\chi)(\bar{q}q)$ and
$(\bar{\chi}\gamma_{\mu}\gamma_5 \chi)(\bar{q} \gamma^{\mu} \gamma_5
q)$, $n=0$ and $i =$ SI, SD, respectively.  Generalizations to these
scenarios include the operator $(\bar{\chi}\gamma_5 \chi)(\bar{q}q)$,
which yields an exponent $n=1$ and a rate that is not dependent on
nuclear spin.  In contrast, $(\bar{\chi} \gamma_5
\chi)(\bar{q}\gamma_5 q)$ has $n=2$ and $i=\text{SD}$.  The momentum
dependence in the rate has an important effect on the recoil spectrum,
suppressing scattering at low energies.  This leads to a peaked recoil
spectrum and potentially more high-energy events than would be
expected for the case of standard elastic scattering with no momentum
dependence, where the rate falls off exponentially.

\subsection{\label{sec:VelocityDist} Astrophysics: Dark Matter Distribution}

The velocity distribution $f(\bv)$ of dark matter particles in the
Galactic halo affects the signal in dark matter detectors.  Here, we
discuss the velocities of the dark matter components of the halo.  The
dominant contribution is a smooth virialized component, discussed in
\refsec{SmoothHalo}.  The formation of the Milky Way via merger events
leads to significant structure in both the spatial and velocity
distribution of the dark matter halo, including dark matter streams
and tidal debris, as discussed in \refsec{UnvirializedHalo}.

Velocity distributions are frequently given in a frame other than the
lab frame to simplify their analytical form.  In this review, we
define $\widetilde{f}(\bv)$ as the distribution in the rest frame of
the dark matter population (\ie\ the frame in which the bulk motion of
the dark matter particles is zero); in the case of the (essentially)
non-rotating smooth halo background, that frame is the Galactic rest
frame.  The lab frame distribution is obtained through a Galilean
transformation as described in \refsec{Modulation}.  More details of
several commonly used distributions, including analytical forms for
the mean inverse speed $\eta$, can be found in \refapp{MIV}.

\subsubsection{\label{sec:SmoothHalo} Smooth Halo Component}

\begin{figure}
  \insertfig{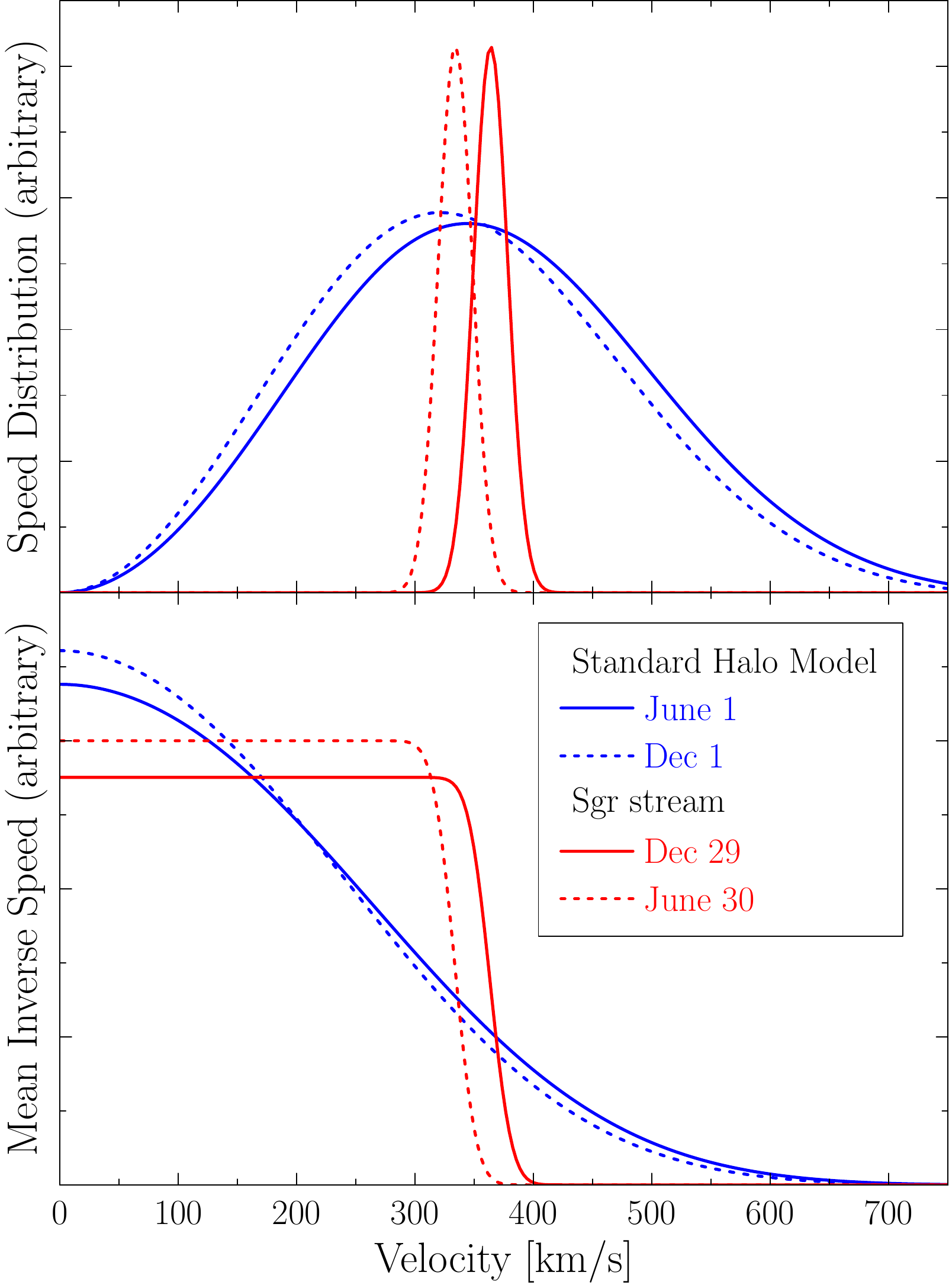}
  \caption{
    Comparison of the Standard Halo Model (SHM) and an example stream,
    representative of the smooth background halo and a cold flow,
    respectively.  The stream, modeled after the Sagittarius (Sgr)
    stream, is roughly orthogonal to the Galactic plane with speed
    $\sim$350 km/s relative to the Sun.  Upper Panel: The speed
    distribution (one dimensional $f(v)$ in the frame of the Earth)
    for both components.  Lower panel: The differential signal in a
    detector is directly proportional to the mean inverse speed
    $\eta(\vmin)$.  Here, the $x$-axis is $\vmin$, the lower limit of
    the integration in \refeqn{eta}.  The approximately exponential
    SHM and step-like stream $\eta$'s are each shown at two periods of
    the year, corresponding to the times of year at which $\eta$ is
    minimized and maximized; note these times are different for the
    two components.
    }
  \label{fig:eta}
\end{figure}

The dark matter halo in the local neighborhood is most likely
dominated by a smooth and well-mixed (virialized) component with an
average density $\rhochi \approx 0.4$~GeV/cm$^3$.\footnote{%
  Estimates for the local density of the smooth dark matter component
  are model dependent and vary in the literature by as much as a
  factor of two \cite{Caldwell:1981rj,Catena:2009mf,Weber:2009pt,
  Salucci:2010qr,Pato:2010yq,Bovy:2012tw}.  Historically, 0.3~GeV/cm$^3$
  has often been assumed when making comparisons between direct
  detection results.  While this density is by no means ruled out by
  current observations, recent estimates tend to suggest a value closer
  to 0.4~GeV/cm$^3$.   Both values of the local density can be found in
  recent direct detection literature.
  }
The simplest model for this smooth component is often taken to be the
Standard Halo Model (SHM)~\cite{Drukier:1986tm,Freese:1987wu}, an
isothermal sphere with an isotropic, Maxwellian velocity distribution
and rms velocity dispersion $\sigma_v$.  The SHM is written as
\begin{equation} \label{eqn:TruncMaxwellian}
  \widetilde{f}(\bv) =
    \begin{cases}
      \frac{1}{\Nesc} \left( \frac{3}{2 \pi \sigma_v^2} \right)^{3/2}
        \, e^{-3\bv^2\!/2\sigma_v^2} , 
        & \textrm{for} \,\, |\bv| < \vesc  \\
      0 , & \textrm{otherwise}.
    \end{cases}
\end{equation}
Here,
\begin{equation} \label{eqn:Nesc}
  \Nesc = \erf(z) - \frac{2}{\sqrt{\pi}} z e^{-z^2} \, ,
\end{equation}
with $z \equiv \vesc/\vmp$, is a normalization factor and
\begin{equation} \label{eqn:vmp}
  \vmp = \sqrt{2/3} \, \sigma_v
\end{equation}
is the most probable speed, with an approximate value of 235~km/s
\cite{Kerr:1986hz,Reid:2009nj,McMillan:2009yr,Bovy:2009dr} (see
\refsec{Modulation} for further discussion).
The Maxwellian distribution is truncated
at the escape velocity $\vesc$ to account for the fact that WIMPs with
sufficiently high velocities escape the Galaxy's potential well and,
thus, the high-velocity tail of the distribution is depleted.  The
dark matter escape velocity in the Milky Way is estimated from that of
high-velocity stars.  The RAVE survey finds that the 90\% confidence
range is 498--608 km/s \cite{Smith:2006ym}.  \Reffig{eta} shows the
SHM speed distribution in the lab (Earth) frame, after accounting for
the motion of the solar system relative to the Galactic rest frame, as
well as the mean inverse speed $\eta$.

The sharp cut-off at the escape speed in \refeqn{TruncMaxwellian} is
not physical.  To smoothen the transition near the escape speed, one
may use the (still ad hoc) distribution:
\begin{equation} \label{eqn:SubMaxwellian}
  \widetilde{f}(\bv) =
    \begin{cases}
      \frac{1}{\Nesc} \left( \frac{3}{2 \pi \sigma_v^2} \right)^{3/2}
        \, \left[ e^{-3\bv^2\!/2\sigma_v^2} - e^{-3\vesc^2\!/2\sigma_v^2} \right] , 
        & \textrm{for} \,\, |\bv| < \vesc  \\
      0 , & \textrm{otherwise} \, ,
    \end{cases}
\end{equation}
where
\begin{equation} \label{eqn:Nesc2}
  \Nesc = \erf(z) - \frac{2}{\sqrt{\pi}} z \, (1+\tfrac{2}{3}z^2) \, e^{-z^2} \, .
\end{equation}
In another approach, Ref.~\cite{Chaudhury:2010hj} uses King models to
obtain the velocity distribution, handling the finite size and mass of
the Galaxy in a more self-consistent manner.  In these models, the
probability distribution can reach zero at a lower velocity than the
escape velocity; essentially, the highest \textit{bound} velocities
are unpopulated.  In general, because of the large uncertainty in
modeling the tail of the velocity distribution, one should approach
any result that depends sensitively on high velocity predictions with
caution.

For the conventional SI and SD elastic scattering, the recoil spectrum
falls off exponentially in the Galactic rest frame for the SHM
(neglecting form factors), due to the exponential drop off with
velocity in \refeqn{TruncMaxwellian}.  Even when form factors and the
motion of the Earth through the halo are accounted for, the spectrum
is still approximately exponential in the lab frame:
\begin{equation} \label{eqn:falloff}
  \dRdEnr \sim e^{-\Enr/E_0} \, ,
\end{equation}
where $E_0$ is some effective scale that is
$\orderof{10~\mathrm{keV}}$ for typical WIMP and nuclear target
masses, so that the largest contribution to the rate in detectors is
at low recoil energies.  For momentum-dependent interaction operators
or inelastic scattering, the rate may instead peak at higher values of
recoil energy.

The isotropic, Maxwellian velocity distribution of
\refeqn{TruncMaxwellian}, intended to describe a class of smooth
spherical halo models, is only a first approximation of the local halo
profile.  As reviewed in Ref.~\cite{Green:2011bv}, oblate/prolate or
triaxial halos would be expected to have an anisotropic velocity
distribution, which may be approximated as
\begin{equation} \label{eqn:anisotropic}
  \widetilde{f}(\bv)
    \propto \exp\left( - \frac{v_{1}^2}{2 \sigma_{1}^2}
                       - \frac{v_{2}^2}{2 \sigma_{2}^2}
                       - \frac{v_{3}^2}{2 \sigma_{3}^2}
                \right) \,, 
\end{equation}
where $v_i$ are the WIMP velocities along three perpendicular
directions with dispersions $\sigma_{i}$.  In general, changes to the
halo shape from anisotropy result in $\orderof{10\%}$ changes in the
annual modulation signal~\cite{Green:2000jg,Green:2010gw}, although a
more exact statement depends on the dark matter properties and the
detector threshold.

High resolution cosmological N-body simulations provide evidence that
a Maxwellian distribution does not fully capture the velocity
distribution of the smooth halo component, particularly along the high
velocity tail, which is important for detection of low mass WIMPs as
detectors are sensitive only to the highest velocity WIMPs in this
case.  Ref.~\cite{Kuhlen:2009vh} determined the velocity distribution
from two of the highest resolution numerical simulations of Galactic
dark matter structure (Via
Lactea~II~\cite{Diemand:2006ik,Diemand:2008in} and
GHALO~\cite{Stadel:2008pn}).  They found more low speed particles than
in a Maxwellian case, and a distribution with a peak that is flatter
in shape.  Alternatively, analytic fits for producing better agreement
with numerical results at the high speed tail have been
obtained~\cite{Lisanti:2010qx,Mao:2012hf}.

Another issue is that most simulations only contain dark matter
particles; simulating baryonic physics is extremely difficult, but
important given that baryons dominate in the inner regions of the
Milky Way.  Gas cooling changes halo shapes from prolate-triaxial to
more spherical when baryons are added \cite{Dubinski:1991bm,
  Kazantzidis:2004vu, Debattista:2007yz, Valluri:2009ir, Zemp:2011},
with velocity distributions that are expected to deviate less from the
standard Maxwellian than those found in dark matter-only simulations.
Ref.~\cite{Purcell:2012sh} studied predictions for dark matter
experiments within the context of an isolated numerical model of a
Milky Way-like system designed to reproduce the basic properties of
the Galaxy by including an equilibrated Galactic stellar disk and the
associated Sagittarius galaxy impact, in addition to dark matter.  The
resulting dark matter velocity distribution still exhibits deviations
from Maxwellian and the calculated recoil spectrum has an increased
number of scattering events at large energies.

Using cosmological simulations, Refs.~\cite{Read:2008fh, Read:2008,
Purcell:2009yp} identified the possibility of a disk-like dark
matter component (``dark disk'') that forms from satellite merger
events.  Ling \etal\ \cite{Ling:2009eh} performed a high-resolution
cosmological N-body simulation with baryons.  They study a Milky Way
sized object at redshift $z=0$ that includes gas, stars and dark
matter to characterize the co-rotating dark disk,
which could play an important role in direct detection experiments
\cite{Bruch:2008rx}.  Equilibrated self-gravitating collisionless
structures have been shown to exhibit Tsallis
distributions~\cite{Tsallis:1987eu, Lima:2002hm, Hansen:2004dg,
Hansen:2005yj}:
\begin{equation} \label{eqn:Tsallis}
  \widetilde{f}(\bv)
    = \frac{1}{N(v_0,q)} \left[ 1-(1-q) \frac{\bv^2}{v_0^2} \right]^{q/(1-q)}
      \, ,
\end{equation}
where $N(v_0,q)$ is a normalization constant and the Maxwell-Botzmann
distribution is recovered by taking the limit $q \rightarrow 1$.
For a spherical shell at the same radial distance as the Sun in the
Ling \etal\ simulation, the velocity distribution is best fit by a
Tsallis distribution with $v_0 = 267.2$~km/s and $q = 0.773$.
In an analysis of the dark matter and stars in the cosmological hydro
simulation of Ref.~\cite{Stinson:2010xe}, the work of
Ref.~\cite{Valluri:2013tj} has also found higher tangential motion in dark
matter particles close to the disk plane than away from it,
consistent with a dark disk.

\subsubsection{\label{sec:UnvirializedHalo} Unvirialized Structure of Halo}

The Milky Way halo forms through the merging of smaller dark matter
subhalos.  These merging events can lead to significant structure in
both the spatial and velocity distribution of the dark matter
halo. High resolution cosmological dark matter simulations, such as
Via Lactea \cite{Diemand:2006ik,Diemand:2008in}, GHALO
\cite{Stadel:2008pn}, and Aquarius \cite{Springel:2008cc}, find
residual substructure from the merging process that includes dark
matter clumps, cold streams, and debris flows.  The dark matter
affiliated with any of these substructures located in the Solar
neighborhood affects count rates and spectra as well as the phase and
amplitude of the annual modulation in experiments
\cite{Gelmini:2000dm, Stiff:2001dq, Freese:2003na, Freese:2003tt,
  Savage:2006qr, Kuhlen:2009vh, Alves:2010pt, Kuhlen:2012fz,
  Purcell:2012sh}.

An example of spatially-localized substructure is a dense clump or
subhalo of dark matter.  If the Earth is sitting in such a clump, the
local dark matter density would be larger than currently expected,
increasing scattering rates in experiments.  According to numerical
simulations, however, local density variations due to the clumpiness
of the dark matter halo are unlikely to significantly affect the
direct detection scattering rate. Based on the Aquarius Project,
Ref.~\cite{Vogelsberger:2008qb} reports that the dark matter density
at the Sun's location differs by less than 15\% from the average at
more than 99.9\% confidence and estimates a probability of 10$^{-4}$
for the Sun being located in a bound subhalo of any mass.  The
possibility that the Earth may reside in a local underdensity due to
unvirialized subhalos throughout the Galaxy should also be taken into
account when interpreting direct detection null results;
Ref.~\cite{Kamionkowski:2008vw} predicts a positively skewed density
distribution with local densities as low as one tenth the mean value,
but probably not much less than half.

In addition to structure in configuration space, the dark matter halo
can also exhibit velocity substructure in the form of debris flows or
cold tidal streams.  Debris flows are an example of
spatially-homogenous velocity substructure that consists of the
overlapping shells, sheets, and plumes formed from the tidal debris of
the (sum total of) subhalos falling into the Milky
Way~\cite{Lisanti:2011as,Kuhlen:2012fz}.  Although this dark matter
component is spatially uniform, the distribution of its Galactocentric
speeds is roughly a delta function.\footnote{%
  Note the distinction between a delta function in the \textit{speed}
  for a debris flow and a delta function in the \textit{velocity} for
  a stream (below).
  }
In Via Lactea~II, more than half of the dark matter near the Sun with
(Earth-frame) speeds greater than 450~km/s is debris flow.  At higher
speeds, debris flow comprises over 80\% of the dark matter.  As a
result, debris flow is particularly important for experiments that
probe the high velocity tail of the dark matter distribution, such as
searches for light dark matter or experiments with directional
sensitivity.

Tidal streams are another unvirialized component of the halo and also
consist of material stripped from infalling satellites.  As the
material in the stream has not had the time to spatially mix, the
stream has a small velocity dispersion in comparison to that of the
virialized halo.  A dark matter stream is coherent in velocity space,
with
\begin{equation} \label{eqn:StreamDist}
  \tilde{f}_{\rm str}(\bv) = \delta^3(\bv)
\end{equation}
in the limit of zero dispersion.  In some cases, particularly when
examining the annual modulation signal, it may be important to account
for the small but non-negligible dispersion of the stream.  In such
cases, a Maxwellian velocity distribution\footnote{%
  Tidal streams can have much more anisotropic velocity distributions
  than the smooth halo background, with a larger dispersion along the
  longitudinal direction than transverse directions
  \cite{Stiff:2001dq}.  Still, the isotropic Maxwellian distribution
  with an appropriately chosen $\sigma_v$ can provide a sufficiently
  good approximation for the purposes of direct detection
  calculations.
  }
can be used, albeit with a much smaller $\sigma_v$ than that for the
SHM. The speed distribution of an example stream in the lab frame, as
well as the corresponding $\eta$, are shown in \reffig{eta}.

The Sagittarius stream is one of the most stunning examples of a
stellar stream in our Galaxy.  Sagittarius (Sgr) is a satellite galaxy
that is located inside the Milky Way on the opposite side of the
Galactic Center from the Sun; it is currently being disrupted and
absorbed by the Milky Way. The Sloan Digital Sky Survey and the Two
Micron All Sky Survey \cite{Yanny:2003zu, Newberg:2003cu,
  Majewski:2003ux} have traced the stellar component of the tidal
stream \cite{Ibata:2000pu, DohmPalmer:2001xp}.  Two streams of matter
are being tidally pulled away from the main body of the Sgr galaxy and
extend outward from it. Whether the Sgr stream passes close enough to
the solar neighborhood to affect direct detection experiments remains
up for debate.  Early data indicated that the leading tail of stellar
material ripped from the Sgr galaxy passes only a few kpc from the
solar neighborhood \cite{Belokurov:2006ms, Seabroke:2007er}, but later
studies indicated that the center of the stream's stellar component
could be farther away than initial estimates suggested
\cite{Fiorentin:2011EAS}.  Most recently, however,
Ref.~\cite{Purcell:2012sh} analyzed self-consistent N-body simulations
of the Milky Way disk and the ongoing disruption of the Sgr dwarf
galaxy and argued that the dark matter part of the Sgr stream may, in
fact, impact the Earth.  Streams can have a variety of effects on
direct detection experiments \cite{Freese:2003tt, Freese:2003na}, as
discussed further below.

Alternative models of halo formation, such as the late-infall model
\cite{Gunn:1972sv, Fillmore:1984wk, Bertschinger:1900nj} more recently
examined by Sikivie and others \cite{Sikivie:1992bk, Sikivie:1996nn,
  Sikivie:1997ng, Sikivie:1999jv, Tremaine:1998nk, Natarajan:2005fh,
  Natarajan:2010jx}, also predict cold flows of dark matter.  In the
caustic ring model \cite{Duffy:2008dk}, the annual modulation is 180
degrees out of phase compared to the usual (isothermal) model.  Any
such streaming of WIMPs (we will henceforth use ``stream'' to imply
any cold flow) will yield a significantly different modulation effect
than that due to a smooth halo.

Finally, there may be unbound dark matter of extragalactic origin
passing through the Galaxy.  If present, these high-speed particles
can increase the number of high-energy scattering events in a direct
detection experiment \cite{Freese:2001hk,Baushev:2012dm}.

\section{\label{sec:Modulation} Annual Modulation}

The velocity distribution in the Earth's frame $f(\bv,t)$ changes
throughout the year due to the time-varying motion of an observer on
Earth.  Assuming $\widetilde{f}(\bv)$ is the velocity distribution in
the rest frame of the dark matter population, \ie\ the frame where the
bulk motion is zero, the velocity distribution in the lab frame is
obtained after a Galilean boost:
\begin{equation} \label{eqn:vdist}
  f(\bv,t) = \widetilde{f}(\bvobs(t) + \bv) \, ,
\end{equation}
where
\begin{equation} \label{eqn:vobs}
  \bvobs(t) = \bv_{\odot} + \bV_{\oplus}(t)
\end{equation}
is the motion of the lab frame relative to the rest frame of the dark
matter, $\bv_{\odot}$ is the motion of the Sun relative to that frame,
and $\bV_{\oplus}(t)$ is the velocity of the Earth relative to the
Sun.  For a non-rotating, smooth background halo component, such as
the SHM, $\bv_{\odot} = \bv_{\mathrm{LSR}} +
\bv_{\odot,\mathrm{pec}}$, where $\bv_{\mathrm{LSR}} = (0,\vrot,0)$ is
the motion of the Local Standard of Rest in Galactic
coordinates,\footnote{%
  Galactic coordinates are aligned such that $\hat{\mathbf{x}}$ is the
  direction to the Galactic center, $\hat{\mathbf{y}}$ is the
  direction of the local disk rotation, and $\hat{\mathbf{z}}$ is
  orthogonal to the plane of the disk.
  }
and $\bv_{\odot,\mathrm{pec}} = (11,12,7)$~km/s is the Sun's peculiar
velocity (see \eg\ Refs.~\cite{Mignard:2000aa, Schoenrich:2009bx} and
references therein).  The canonical value for the disk rotation speed
$\vrot$ has long been 220~km/s \cite{Kerr:1986hz}, but more recent
estimates tend to place it 5--15\% higher \cite{Reid:2009nj,
McMillan:2009yr, Bovy:2009dr}.  A value of 235~km/s is more centrally
located within current estimates and is more frequently being used as
a fiducial value, though 220~km/s remains viable.

The $\bV_{\oplus}(t)$ term in \refeqn{vobs} varies throughout the year
as the Earth orbits the Sun, leading to an annual modulation in the
velocity distribution and, thus, the recoil rate.  Written out in
full,
\begin{equation} \label{eqn:voplus}
  \textbf{V}_{\oplus}(t) = 
              V_\oplus \left[
                  \eone \cos{\omega(t-t_1)} + \etwo \sin{\omega(t-t_1)}
                \right] \, ,
\end{equation}
where $\omega = 2\pi$/year, $V_\oplus = 29.8$ km/s is the Earth's
orbital speed around the Sun, and $\eone$ and $\etwo$ are the
directions of the Earth's velocity at times $t_1$ and $t_1+0.25$
years, respectively.  \Refeqn{voplus} neglects the ellipticity of the
Earth's orbit, which is small and only gives negligible changes to the
velocity expression (see Refs.~\cite{Green:2003yh,Lewin:1995rx} for
more detailed expressions).  In Galactic coordinates,
\begin{equation}
  \label{eqn:eone}
    \eone = (0.9931, 0.1170, -0.01032)
    \quad \text{and} \quad
    \etwo = (-0.0670, 0.4927, -0.8676) \, ,
\end{equation}
where $\eone$ and $\etwo$ are the directions of the Earth's motion at
the Spring equinox (March 21, or $t_1$) and Summer solstice (June 21),
respectively.

If we define the characteristic time $t_0$ as the time of year at
which $\vobs(t)$ is maximized, \ie\ the time of year at which Earth is
moving fastest with respect to the rest frame of the dark matter, then
the magnitude of $\vobs(t)$ is approximately
\begin{equation} \label{eqn:vobst}
  \vobs(t) \approx v_{\odot} + b V_\oplus \cos{\omega(t-t_0)} ,
\end{equation}
where $b \equiv \sqrt{b_1^2 + b_2^2}$ for $b_i \equiv \egen \cdot
\hat{\bv}_\odot$ is a geometrical factor associated with the direction
of $\bvobs$ relative to Earth's orbital plane (note $|b| \le 1$).  The
approximation is valid when $V_{\oplus}/v_{\odot} \ll 1$, as is the
case with nearly all halo components.

Once the Galilean transformation of \refeqn{vdist} is performed,
$\eta(\vmin,t)$ is calculated via \refeqn{eta}; see \refapp{MIV} for
analytical forms of $\eta$ for several commonly used distributions.
Because the modulation rate must have a fixed period of one
year,\footnote{%
  The density and intrinsic velocity distribution will change as the
  solar system moves into, through, and back out of any substructure
  of finite size, such as a clump, leading to variations in the recoil
  rate that do not manifest as an annual modulation.  However, the
  time scales involved are typically many orders of magnitude longer
  than a year and such temporal variations can be neglected.
  }
the differential scattering rate can be expanded in a Fourier series:
\begin{equation} \label{eqn:Fourier}
\dRdE(\vmin,t)
    = A_0 + \sum_{n=1}^{\infty} A_n \cos{n\omega(t-t_0)}
          + \sum_{n=1}^{\infty} B_n \sin{n\omega(t-t_0)} \, ,
\end{equation}
where the Fourier coefficients $A_n$ and $B_n$ are functions of
$\vmin$ (see~\refeqn{vmin}).  If the velocity distribution in the rest frame of the dark
matter is isotropic, then $B_n = 0$.  This simplification is a direct
result of expanding about $t_0$: though the Fourier expansion could be
made about any other (arbitrary) phase, the sum would include both
cosine and sine terms.  The expansion in terms of only cosines should
not be surprising as $\vobs(t)$ contains only a single cosine term in
\refeqn{vobst}.  This simplification does not apply for anisotropic
distributions; however, for nearly all realistic anisotropic
distributions, $B_n \ll A_n$, and the sine terms in the expansion can
still be neglected.  As a consequence, the modulation will always be
symmetric (or very nearly so) about the characteristic time $t_0$.

For smooth components of the halo, such that $f(\bv)$ is slowly
varying over $\delta\bv \sim V_{\oplus}$, one further finds that $A_0
\gg A_1 \gg A_{n\ge2}$, assuming $v_{\odot} \gg V_{\oplus}$ as is the
case for most components of the halo.  This relation further holds for
any structure in the halo when $f(\bv)$ is slowly varying for $|\bv|
\approx \vmin$, as is the case for cold flows where $\vmin \not\approx
v_{\odot}$.  Higher order terms in the Fourier expansion may become
important when $f(\bv)$ exhibits sharp changes in the vicinity of
$|\bv| \approx \vmin$, which can happen in the case of a stream.

Except for the special cases described above, the annually modulating
recoil rate can be approximated by
\begin{equation} \label{eqn:dRdES}
  \dRdE(E,t) \approx S_0(E) + S_m(E) \cos{\omega(t-t_0)} ,
\end{equation}
with $|S_m| \ll S_0$, where $S_0$ is the time-averaged rate, $S_m$ is
referred to as the modulation amplitude (which may, in fact, be
negative), $\omega = 2\pi$/year and $t_0$ is the phase of the
modulation.\footnote{%
  Experiments may quote the average amplitude over some interval,
  \begin{equation} \label{eqn:AverageSm}
    \bar{S}_m = \frac{1}{E_2 - E_1} \int_{E_1}^{E_2} \rmd E \, S_m(E) .
  \end{equation}
  }
The quantities $S_0$ and $S_m$ correspond to $A_0$ and $A_1$,
respectively, in the Fourier expansion of \refeqn{Fourier}, but the
former are the standard notation in the literature when only the
constant and first cosine term of the Fourier expansion are used.

In addition to the time-varying motion of a detector due to the orbit
of the Earth about the Sun, there is a time-varying motion due to the
rotation of the Earth about its axis, leading to a daily (diurnal)
modulation in the recoil rate.  This modulation can be determined by
repeating the above procedure with the inclusion of this rotational
velocity term in \refeqn{vobs}.  However, the rotational velocity (at
most 0.5~km/s, near the equator) is significantly smaller than the
orbital velocity (30~km/s), making the daily modulation signal much
smaller than the annual modulation signal and, unfortunately, much
more difficult to detect (made further difficult by the statistical
issues of extracting the modulation from an experimental result, as
discussed below).  For this reason, the daily modulation in the recoil
rate is typically ignored in modulation searches.
A related, but different, effect is the daily modulation in the recoil
\textit{direction}, a much larger effect that may be observed by
directional detectors, to be briefly discussed in \refsec{Future}.

Detecting the modulation signal in an experiment is made 
difficult by the fact that the modulation $S_m$ must be extracted from
on top of a large constant rate $S_0$.  Here, we use a very simple
two bin analysis to illustrate the statistical issues in experimentally
extracting a modulation amplitude.  Suppose an experiment counts events
over a one year period, dividing those events into the six month periods
centered on $t_0$ and $t_0+0.5$~years; we will use `$+$' and `$-$'
subscripts to refer to these two respective periods.  Assuming a
modulation of the form given by \refeqn{dRdES}, these are the periods
when the rate is above and below average, respectively.  The experimental
estimates of the average rate and the modulation amplitude are
\begin{equation} \label{eqn:2binS}
  \Scp \sim \frac{1}{MT\,\Delta E} \, (N_{+} + N_{-})
  \qquad\text{and}\qquad
  \Smp \sim \frac{1}{MT\,\Delta E} \, (N_{+} - N_{-}) \, ,
\end{equation}
where $MT\,\Delta E$ is the exposure of the detector, $M$ is the target
mass, $T=1$~year is the total exposure time, $\Delta E$ is the width of
the energy range considered, and $N_{\pm}$ are the number of events
measured in each bin.  The uncertainty $\delta S_m$ in the amplitude can be
determined from simple error propagation in terms of the two
measurements $N_{\pm}$:
\begin{equation} \label{eqn:2binSmerr}
  (\delta \Smp)^2
    \ =\ \left(\frac{\partial\Smp}{\partial N_{+}} \right)^2 (\delta N_{+})^2
       + \left(\frac{\partial\Smp}{\partial N_{-}} \right)^2 (\delta N_{-})^2
    \ \sim\ \left(\frac{1}{MT\,\Delta E} \right)^2 (N_{+} + N_{-}) \, ,
\end{equation}
where $\delta N_{\pm} = \sqrt{N_{\pm}}$ are the errors in the counts.
The statistical significance of the measured modulation amplitude is
\begin{equation} \label{eqn:2binsignificance}
  \frac{\Smp}{\delta \Smp}
    \ \propto\ \sqrt{\frac{MT\,\Delta E \, \Smpsq}{\Scp}}
    \ \propto\ \sqrt{N_T} \; \frac{\Smp}{\Scp},
\end{equation}
where $N_T \equiv N_{+}+N_{-}$ is the total number of events.  While
this derivation is for a simple two bin analysis of the yearly
modulation, the above proportionality relationship holds true for any
modulation signal and analysis scheme: a reduction in the modulation
amplitude $S_m$ by a factor of 2 would require an increase in the
number of detected events $N_T$ (and hence exposure) by a factor of 4
to be detected to the same statistical significance.
Thus, to detect the daily modulation signal to the same significance
as the annual modulation signal, where the amplitude of the former
is $\gae$60 times smaller than the latter (Earth's surface rotational
speed of $\lae$0.5~km/s versus an orbital speed of 30~km/s), would
require an increase in exposure by a factor of at least $\orderof{60^2}$,
a daunting task.

\begin{figure*}
  \insertwidefig{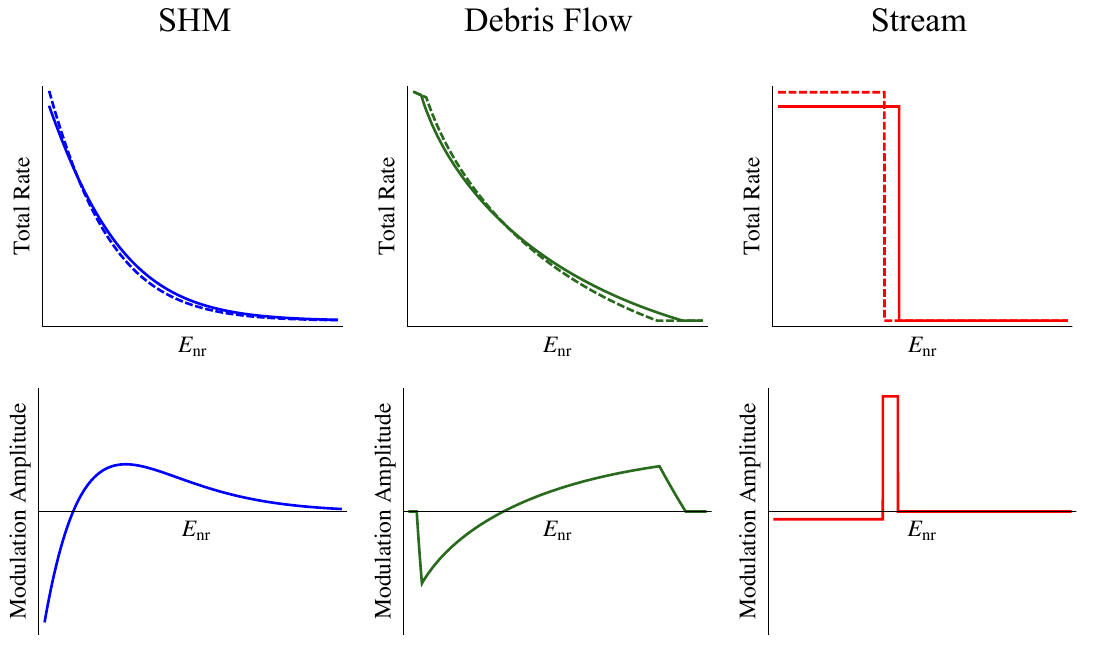}
  \caption{
    A comparison of the shapes of the total rate shown at two periods
    of the year, corresponding to the times of year at which the rate
    is minimized and maximized, as well as the modulation amplitude,
    for three different halo components: SHM (left), debris flow
    (middle), stream (right).  The normalization between panels is
    arbitrary.
    }
  \label{fig:schematic}
\end{figure*}

In the remainder of this section, we examine the modulation for the
SHM and substructure components.  \Reffig{schematic} summarizes the
conclusions we reach.  Note that the expected modulation amplitude
depends sensitively on the assumed dark matter velocity distribution.
In reality, the local dark matter is likely comprised of both a
virialized and unvirialized component, meaning that a signal at a
direct detection experiment may be due to several different dark
matter components.  In this case, a modulation of the form given by
\refeqn{dRdES} with a fixed phase $t_0$ may not be a good
approximation; the shape of the modulation for the total rate may no
longer be sinusoidal in shape and/or the phase may vary with $\vmin$.
Furthermore, there are cases when \refeqn{dRdES} is a bad
approximation even for a single halo component; an example will be
shown below for a stream.  We close this section with a discussion of
what can be learned about the local halo in these more complicated
scenarios.

\subsection{\label{sec:SHM} Smooth Background Halo:
                   Isothermal (Standard) Halo Model}

We now apply our general discussion of the modulation rate to the
example of a simple isothermal sphere~\cite{Freese:1987wu}.  As
discussed in the previous section, the SHM is almost certainly not an
accurate model for the dark matter velocity distribution in the Milky
Way.  However, its simple analytic form provides a useful starting
point for gaining intuition about the modulation spectrum of the
virialized dark matter component.

As we showed in \refeqn{dRdEnr2}, the differential count rate in a
detector is directly proportional to the mean inverse speed $\eta$;
the time-dependence of the recoil rate arises entirely through this
term.  To study the expected time-dependence of the signal in the
detector, we therefore focus on the time-dependence of $\eta$; in
particular, we can investigate the annual modulation of the quantity
$\eta$ as it is the same as that of the dark matter count rate.

For the SHM or any dark matter component with a velocity distribution
described by \refeqn{TruncMaxwellian} or \refeqn{SubMaxwellian}, the
mean inverse speed has an analytical form, presented in \refapp{MIV}
and in Refs.~\cite{Savage:2006qr,McCabe:2010zh}.  \Reffig{eta}
illustrates $\eta(\vmin)$ for the SHM, taking $\vmp = \vrot$ as
expected for an isothermal spherical halo.

\Reffig{eta} shows $\eta(\vmin)$ at $t_0 \simeq$ June~1, the time of
year at which Earth is moving fastest through the SHM, as well as on
December~1, when the Earth is moving slowest; there is a (small)
change in $\eta$ over the year.  The corresponding recoil spectra, as
a function of recoil energy, are given in schematic form in the first
panel of \reffig{schematic}.  The amplitude of the modulation,
\begin{equation} \label{eqn:SmSHM}
  A_1(E) \approx
    \frac{1}{2} \left[
      \dRdE(E,\,\textrm{June 1}) - \dRdE(E,\,\textrm{Dec 1})
    \right] ,
\end{equation}
is also shown in the figure.  Two features of the modulation are
apparent for the SHM: (1) the amplitude of the modulation is small
relative to the average rate, with an exception to be discussed below,
and (2) the amplitude of the modulation changes sign at small $\vmin$
(low recoil energies).  This phase reversal can be used to constrain
the WIMP mass.

\begin{figure}
  \insertfig{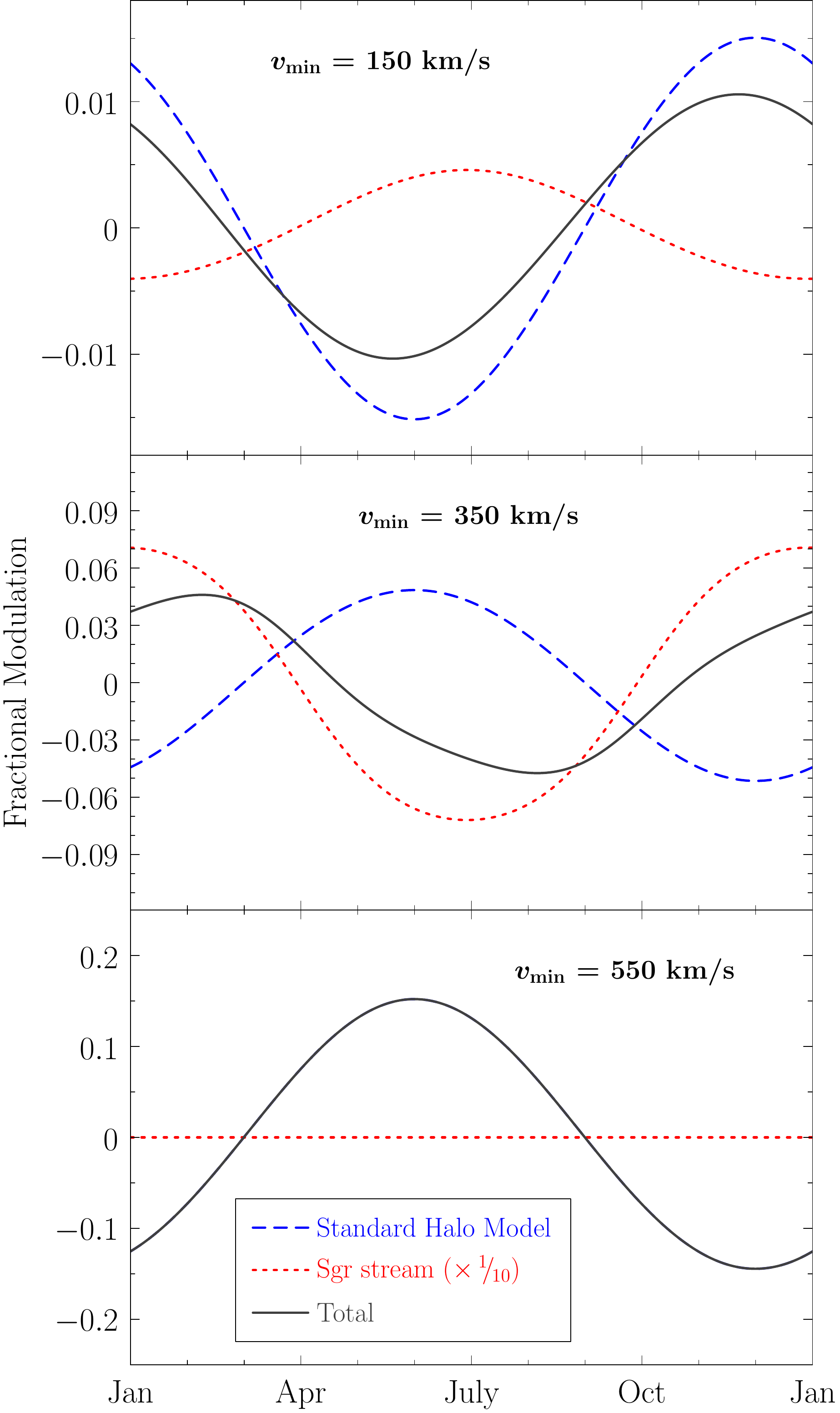}
  \caption{
    The residual rate for the SHM (dashed) and an example stream
    (dotted) is plotted at several values of $\vmin$.  The stream is
    modeled after the Sgr stream.  Also shown is the total SHM+stream
    modulation, assuming the local density of the stream is 10\% that
    of the SHM.  The residual rates are given relative to the average
    rate in each case, \ie\ curves show the \textit{fractional}
    modulation, except for the stream, where the relative rate has
    been divided by ten for visual clarity as its relative modulation
    is much larger.  The corresponding recoil energy for each $\vmin$,
    given by \refeqn{vmin}, depends on the WIMP and nuclear target
    masses; for a WIMP mass of 60~GeV and a germanium target, the
    $\vmin$ values of 150, 350, and 550~km/s correspond to recoil
    energies of 7, 40, and 100~keV, respectively.  The phase reversal
    of the SHM component, which occurs at smaller $\vmin$, can be seen
    by comparing the dashed blue curves in the top two panels.  The
    recoil energy at which this phase reversal takes place can be used
    to determine the WIMP mass \cite{Lewis:2003bv}.
    }
  \label{fig:modulation}
\end{figure}

\Reffig{modulation} illustrates the residual time-varying signal for
the SHM (dashed blue curves). The different panels show how the
modulation depends on $\vmin$.  In general, the modulation has a
sinusoidal shape and is symmetric about $t_0$; the sinusoidal shape
allows for the use of the amplitude approximation given by
\refeqn{SmSHM}.  For small $\vmin$ (low recoil energies), the rate is
\textit{minimized} at a time $t_0$, while for large $\vmin$ (high
recoil energies), the rate is \textit{maximized} at $t_0$.

An important quantity of interest is the \textit{modulation fraction},
defined as the size of the modulation amplitude relative to the
average total rate: $S_m/S_0$.  For a wide range of $\vmin$, the
modulation fraction is $\orderof{1-10\%}$, as seen in
\reffig{modulation}.  For $\vmin$ above $\sim 500$~km/s, both $S_0$
and $S_m$ fall rapidly with $\vmin$ as scatters come only from WIMPs
in the tail of the Maxwellian distribution.  In this region, $S_0$
falls more rapidly, so the modulation fraction grows, going from
$\orderof{10\%}$ to $\orderof{100\%}$.  Because of the low absolute
rates at these higher energies, experiments are generally not
sensitive to the $\vmin$ region with high modulation fraction.
However, for WIMPs that are much lighter than the nuclear target,
large modulation fractions do correspond to the recoil energies of
interest in detectors, and an order unity modulation can be
observed.\footnote{%
  When the size of the variations in the recoil rate throughout the
  year becomes comparable to the average rate, \ie\ the relative
  modulation amplitude is of order unity, \refeqn{SmSHM} is no longer
  a good approximation to the modulation.  A significant deviation
  from a cosinusoidal modulation would be an expected signature for
  large modulation fractions.
  }

As noted above, the phase reversal of the annual modulation,
illustrated in \reffig{modulation}, can be used to determine the WIMP
mass \cite{Lewis:2003bv} and is perhaps the best feature of a direct
detection signal for doing so.  While the phase of the modulation is
fixed for a given $\vmin$, regardless of the WIMP mass, the phase of
the modulation for a given recoil energy $\Enr$ is not, as the $\Enr$
associated with a given $\vmin$ is dependent on the WIMP mass through
\refeqn{vmin}.  Thus, an experimental determination of the recoil
energy at which the phase reverses, which occurs at $\vmin \approx
210$~km/s for the SHM, can constrain the WIMP mass.  For a germanium
detector, the SHM phase reversal is expected to occur at recoil
energies of 1, 5, and 15~keV for WIMP masses of 10, 25, and 60~GeV,
respectively; the modulation spectrum should be readily
distinguishable between these cases.  As the WIMP gets much heavier
than the target nucleus, the recoil spectrum becomes degenerate and
the energy of the phase reversal approaches a fixed value (62~keV for
germanium); observation of a reversal at this energy allows only a
lower limit to be placed on the WIMP mass.\footnote{%
  A caveat regarding extracting limits on the WIMP mass from the phase
  reversal is in order.  As we will illustrate shortly, cold flows
  (such as streams or caustics) can strongly affect the phase of the
  modulation.  Thus, the phase of the annual modulation constrains the
  WIMP mass only when the distribution of particle velocities in the
  solar neighborhood is known. One may however use the results of two
  different experiments to constrain the mass without assuming a form
  for the velocity distribution \cite{Drees:2007hr,Drees:2008bv}.
  }
We emphasize that detection of this phase reversal could constitute an
important signature of a WIMP flux, as backgrounds would not give rise
to such an effect.

As discussed in the previous section, the SHM may not be an accurate
model of the smooth background halo.  However, the generic features of
the SHM modulation signal discussed here are also features to be
expected of any smooth background halo model.  In particular, the
modulation should be sinusoidal in shape with a phase around the
beginning of June, have an $\orderof{1-10\%}$ modulation amplitude
(except at high $\vmin$), and have a phase
reversal~\cite{Green:2000jg, Green:2002ht, Green:2010gw}.  However,
all these features can be significantly altered if there is any
significant substructure present in the halo, so we turn to
substructure next.

\subsection{\label{sec:SgrStream} Halo Substructure}

Next, we consider the modulation spectrum when the dark matter
scattering in the detector is dominated by unvirialized components in
the halo, such as debris flows or streams.  The schematic in
\reffig{schematic} illustrates how the modulated and total rates in
these cases can differ drastically from that expected from a smooth
halo contribution.  For a complete discussion of the modulation
spectra for dark matter streams, see Ref.~\cite{Savage:2006qr}.  A
simple analytic approximation for the mean inverse speed of the debris
flow is given in~\cite{Kuhlen:2012fz}, from which it is
straightforward to derive the modulated and total rates (see also
\refapp{MIV}).

\Reffig{schematic} emphasizes that substructure components can
increase the number of expected scattering events at large recoil
energies, in comparison to the smooth halo contribution.  This is due
to the fact that velocity substructure is most likely to be found near
the escape velocity, where the dark matter is predominantly not in
equilibrium.  In particular, while the modulation amplitude for a
smooth halo falls with energy, that for debris flows and streams can
peak at large values.  Therefore, a larger-than-expected modulation at
high recoil energies can be an important indicator for dark matter
substructure in the local neighborhood.

As an example of a substructure component, we now consider the case of
a dark matter stream more fully.  For the case of a dispersionless
dark matter stream, the recoil spectrum is proportional to
\begin{equation} \label{eqn:eta0}
  \eta_{\mathrm{str}}(\vmin,t) = \frac{\theta(\vobs(t) - \vmin)}{\vobs(t)} \, ,
\end{equation}
where $\theta$ is the Heaviside function, and is flat up to the cutoff
energy
\begin{equation} \label{eqn:Ec}
  E_{c}(t) = \frac{2 \mu^2}{M} \vobs(t)^2 \, .
\end{equation}
This characteristic energy is the maximum recoil energy that can be
imparted to the nucleus, and is obtained as follows: The maximum
momentum transferred from a WIMP to a nucleus occurs when the nucleus
recoils straight back and is $\qmax = 2 \, \mu \, \vobs(t)$.  The
maximum recoil energy of the nucleus then follows as $E_c(t) =
\qmax^2/(2 M)$.  A small, but non-zero velocity dispersion $\sigma_v$
expected in \eg\ tidal streams can soften the sharp edge of the
step-shaped $\eta$.  The velocity dispersion for the Sagittarius
stellar stream, for example, is roughly $\mathcal{O}$(20~km/s)
\cite{Yanny:2003zu,DohmPalmer:2001xp,Majewski:2003ux,Carlin:2011bc}.
The dark matter in a tidal stream can be expected to have a velocity
dispersion of a similar magnitude, though how closely it matches the
dispersion of the stars remains an open question.

We take as an example a stream with velocity and direction similar to
what may be expected if the Sgr stream is accompanied by a broader
stream of dark matter that passes near the solar
neighborhood~\cite{Freese:2003na, Freese:2003tt, Purcell:2012sh}.
This is intended as a concrete example of a more general phenomena,
and will illustrate the basic features of an annual modulation signal
in the presence of a stream.  The stream in our example is roughly
orthogonal to the Galactic Plane and moves at a speed $\sim$350~km/s
relative to the Sun.  Its speed distribution and mean inverse speed
$\eta$ are shown in \reffig{eta}; in the latter case, the step-like
spectrum (with a softened edge) is evident.  For this stream,
$\vobs(t)$ is maximal on December~29 and minimal on June~30.

There are two distinct features of a stream's recoil spectrum that
modulate: (1) the height of the step and (2) the location of the step
edge.  Unlike the SHM, the relative modulation amplitude is fairly
uniform at all $\vmin$ below the velocity of the stream in the lab
frame, though, like the SHM, the modulation is small compared to the
total rate.  This modulation, seen in the top panel of
\reffig{modulation}, is sinusoidal and peaks in late December.  Above
the stream's velocity in the lab frame, the signal vanishes because
there are no available dark matter particles (lower panel of
\reffig{modulation}).

The modulation becomes interesting when the minimum scattering
velocity is approximately equal to the stream's velocity in the lab
frame, $\vmin \approx v_{\odot} \approx$~350~km/s.  This occurs near
the edge of the step in the recoil spectrum.  As illustrated in
\reffig{schematic}, $\eta$ changes by a relatively large amount
throughout the year for $\vmin \approx v_{\odot}$, leading to a very
large modulation.  The modulation at the softened edge of the example
stream, shown in the middle panel of \reffig{modulation}, has a
relative amplitude of nearly 70\%, far larger than that for the SHM.
Though not apparent in this figure, this is one of the special cases
where the higher order terms of \refeqn{Fourier} can become important
and the modulation deviates from a sinusoidal shape; see Figure~7 of
Ref.~\cite{Savage:2006qr} for a clearer example.  This very large
modulation, which occurs only over a narrow range of recoil energies,
has a phase reversed from that at lower energies.

The features of the modulation for our example stream are expected of
any cold flow (stream).  Up to some cutoff energy, the modulation is
uniform, relatively small, and sinusoidal.  Above the cutoff energy,
the modulation, as well as the total rate, is negligible.  Over a
narrow range of energies about the cutoff energy, the modulation can
be very large and possibly non-sinusoidal.  However, the phase of the
modulation and cutoff energy can vary significantly depending on the
direction and speed of the stream.  Observation of unexpected phases
in the modulation and/or a narrow energy range containing an unusual
modulation behavior would not only indicate that a significant stream
of dark matter is passing through the local neighborhood, but would
allow the direction and/or speed of that stream to be determined.
However, there may be more than one significant stream or other
substructure---in addition to the smooth halo background---so more
than one halo component may make significant contributions to the
recoil spectrum and modulation signals, complicating the
interpretation of modulation data.  We turn to a multiple component
halo next.

\subsection{\label{sec:Components} Multiple Component Halo}

Thus far, we have considered the modulation spectra of individual
components of the dark matter halo separately.  However, in greater
likelihood, the local dark matter will be comprised of a combination
of virialized and unvirialized components.  In such cases,
low-velocity dark matter will most likely be in equilibrium and
well-described by a smooth halo, while the high-velocity tail of the
distribution may have additional contributions from streams or debris
flows.  The resulting modulation spectrum will be a linear combination
of the spectra shown in \reffig{schematic}, appropriately weighted by
the relative density of each component.  A general study of dark
matter detection in the presence of arbitrary streams or debris flow,
in combination with a smooth halo background, can be found in
Refs.~\cite{Savage:2006qr,Kuhlen:2012fz}.

The addition of even a small amount of substructure to the smooth halo
background can significantly alter the observational signals in direct
detection experiments from those due to the background distribution
alone.  We take, for example, the case of the SHM with the addition of
our example (Sgr-motivated) stream at a density 10\% that of the SHM.
While the overall recoil spectrum is approximately exponentially
falling due to the large contribution from the SHM, a noticeable
drop-off in the spectrum appears around a characteristic energy $E_c$
corresponding to the step in the stream contribution.  The impact on
the modulation, shown in \reffig{modulation}, is even more pronounced.
As can be seen in the middle panel, the shape can differ significantly
from that due to either contribution alone: the modulation is no
longer sinusoidal and is not even symmetric in time.  The phase also
differs significantly: the peak of the modulation occurs at a time
several months different from that of either component.  From the
different panels, it is also clear that the phase changes with recoil
energy by more than just a 180$^{\circ}$ phase flip.

In the general case where multiple components contribute significantly
to the scattering, the following features in the modulation spectrum
can arise:
\begin{itemize}
  \item The phase of modulation can vary strongly with recoil energy
        and not just by a 180$^{\circ}$ phase reversal;
  \item The combined modulation may not be sinusoidal, even if the
        modulation of each individual component is;
  \item The combined modulation may not be time-symmetric, even if the
        modulation of each individual component is; and
  \item The minimum and maximum recoil rates do not necessarily occur
        0.5~years apart.
\end{itemize}
More quantitatively, the time-dependence of the rate is no longer
dominated by the $A_1$ term in the Fourier expansion of
\refeqn{Fourier}, and other terms in the expansion contribute.  A
power spectrum of the modulation is very useful for understanding the
relative strengths of these higher-order contributions (see \eg\
Ref.~\cite{Chang:2011eb}).  The DAMA experiment is currently the only
one with enough data to have produced a power spectrum of their
results; their measured limit on $A_2/A_1$ can already provide
constraints on certain types of streams, as has been shown for the
case of inelastic dark matter in Ref.~\cite{Alves:2010pt}.

\section{\label{sec:Experiments} Experimental Status of Annual Modulation}

In this section, we discuss the experimental status of dark
matter annual modulation searches.  An extremely diverse set of direct
detection experiments exists, which take advantage of a variety of
target materials and background rejection techniques.  The advantage
of such diversity is that different targets are more or less sensitive
to different types of dark matter and/or features in the velocity
profile.  For example, searches for spin-dependent interactions
require the use of targets with non-zero spin.  Also, a lighter
target, such as germanium or sodium versus xenon, is better for
detecting light mass dark matter.

The current anomalies from DAMA, CoGeNT, and CRESST have engendered a
great deal of excitement in the field, with debates as to whether they
represent the first direct observation of dark matter.  We now review
these experiments, as well as their counterparts that report the
tightest constraints.  We caution that the experimental situation is
rapidly changing; the reader should consult more recent literature for
the current status of the field.

\subsection{\label{sec:Current} Experiments and Results}

\subsubsection{\label{sec:DAMA}{The DAMA Experiment}}

\begin{figure*}
  \insertwidefig{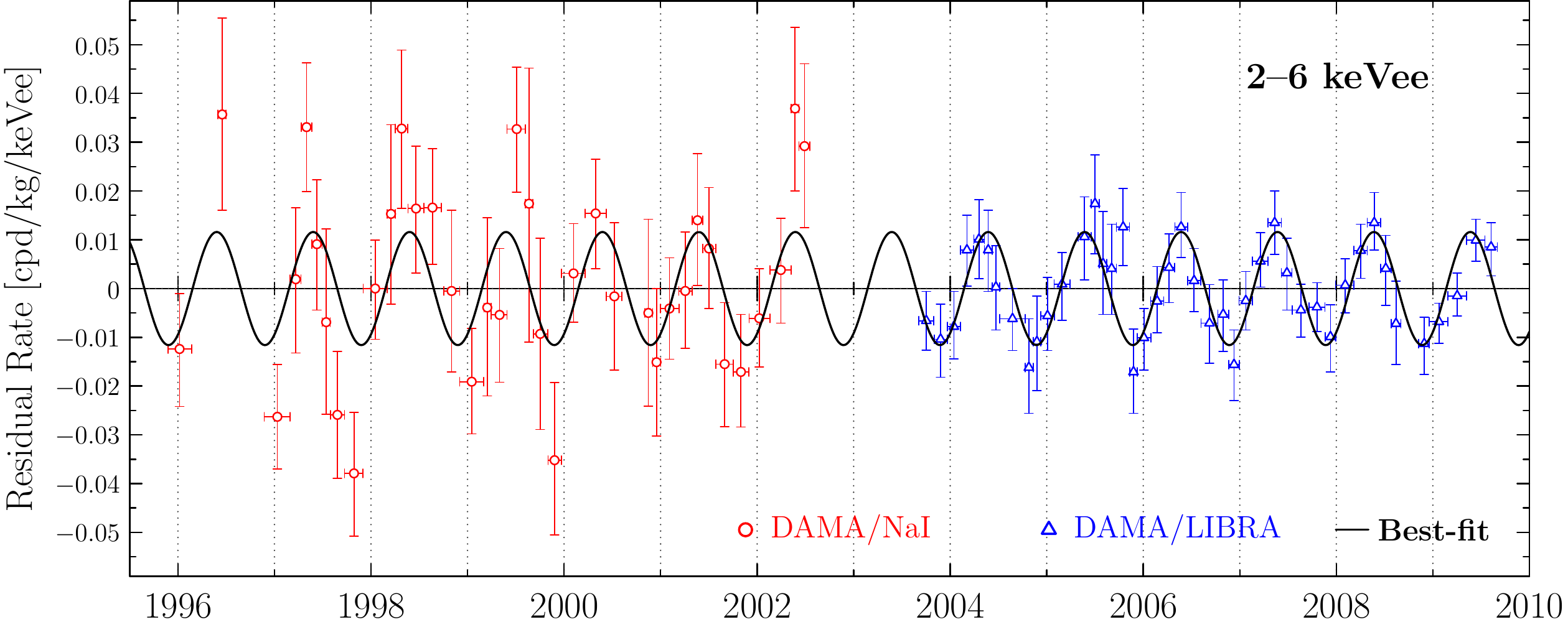}
  \caption{
    The residual rate measured by DAMA/NaI (red circles, 0.29~ton-yr
    exposure over 1995--2002) and DAMA/LIBRA (blue triangles,
    0.87~ton-yr exposure over 2003--2010) in the 2--6~keVee energy
    interval, as a function of time.  Data is taken from
    Refs.~\cite{Bernabei:2003za,Bernabei:2010mq}.  The solid black
    line is the best fit sinusoidal modulation $A
    \cos[\frac{2\pi}{T}(t-t_0)]$ with an amplitude $A = 0.0116\pm
    0.0013$~cpd/kg/keV, a phase $t_0 = 0.400 \pm 0.019$~yr (May~26
    $\pm$ 7~days), and a period $T = 0.999 \pm 0.002$~yr
    \cite{Bernabei:2010mq}.  The data are consistent with the SHM
    expected phase of June~1.
    }
  \label{fig:DAMA}
\end{figure*}

\begin{figure}
   \insertfig{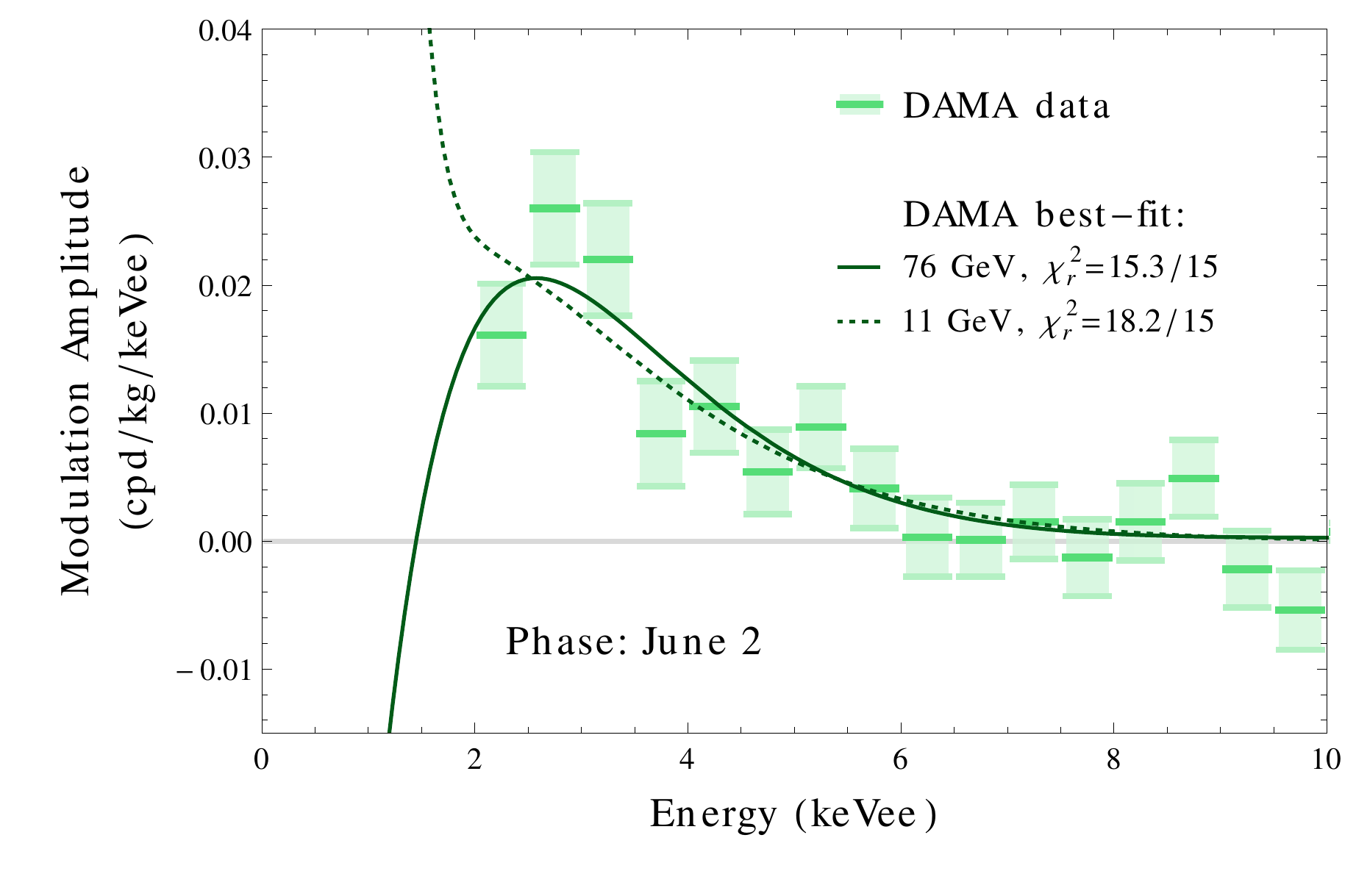}\\
   \insertfig{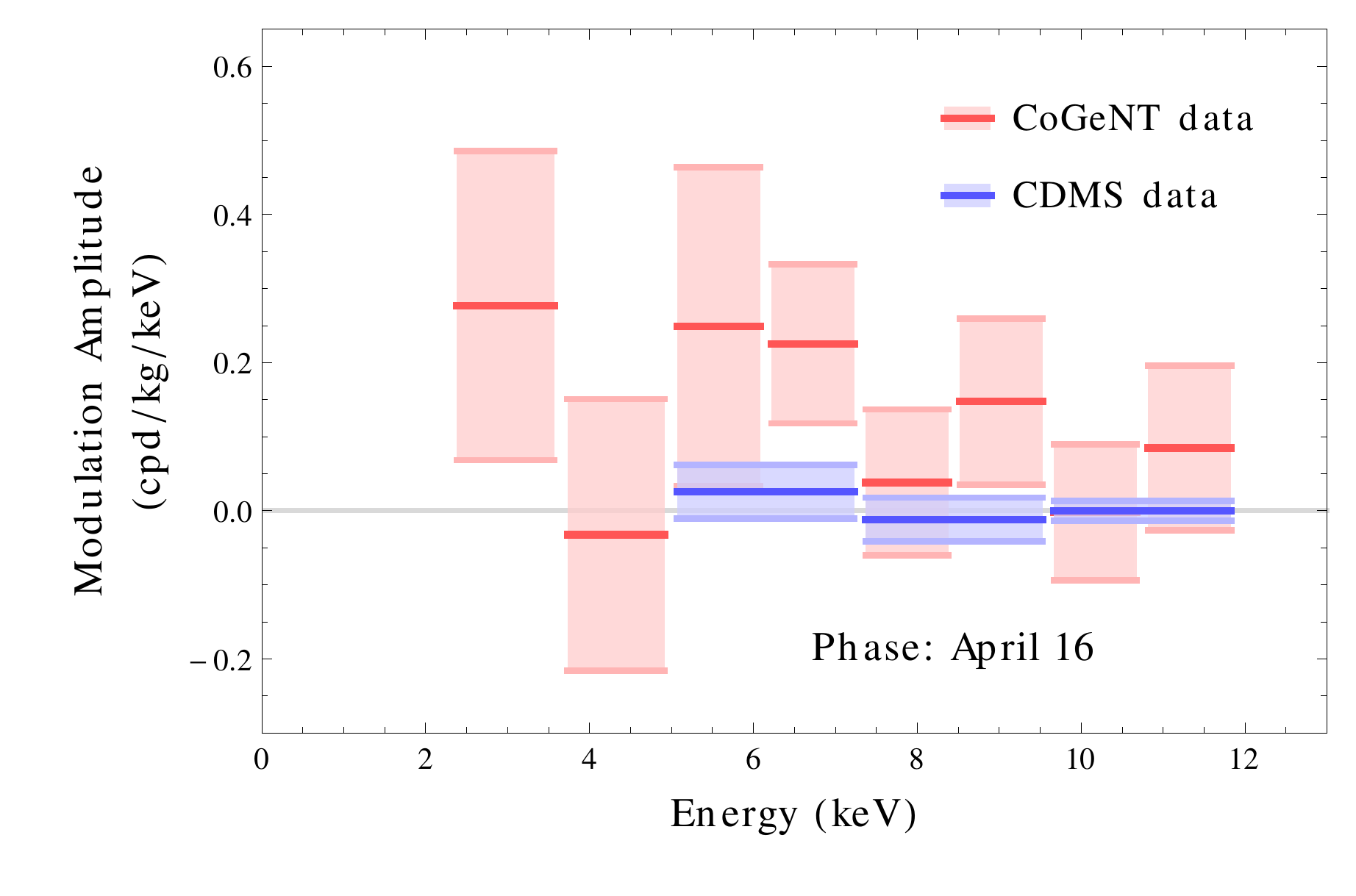}
  \caption{
    The bin-averaged modulation amplitude observed by DAMA (top),
    CoGeNT (bottom, red), and CDMS (bottom, blue) as a function of
    energy.  Boxes indicate the 1$\sigma$ uncertainty for each bin.
    The DAMA results are fit at the SHM expected phase with a peak on
    June~1, while the CoGeNT and CDMS bins are given at the CoGeNT
    best-fit phase with a peak on April~16.  The DAMA data are from
    Ref.~\cite{Bernabei:2010mq} while the CoGeNT and CDMS binning and
    results are taken from Ref.~\cite{Ahmed:2012vq}.  To allow for
    direct comparison between the two germanium experiments, we
    present both CoGeNT and CDMS data in keV (nuclear recoil energy);
    on the other hand, DAMA data are presented in keVee
    (electron-equivalent energy).  Also shown for DAMA are the
    best-fit spectra to the data for spin-independent (SI) scattering,
    corresponding to a WIMP with mass 11~GeV (76~GeV) and SI
    cross-section 2$\times$$10^{-4}$~pb (1.5$\times$$10^{-5}$~pb).
    }
  \label{fig:data}
\end{figure}

The Italian Dark Matter Experiment (DAMA) consists of 250~kg of radio
pure NaI(Tl) scintillator.  DAMA/NaI \cite{Bernabei:2003za} was the
first experiment to claim a positive dark matter signal; it was later
replaced by DAMA/LIBRA \cite{Bernabei:2008yh,Bernabei:2010mq}, which
confirmed the results. The experiment has now accumulated 1.17~ton-yr
of data over 13 years of operation and claims an 8.9$\sigma$ annual
modulation with a phase of $\text{May 26}\pm7$ days, consistent with the
dark matter expectation (see \reffig{DAMA}).  The modulation amplitude
from 2--10~keVee, taken from Ref.~\cite{Bernabei:2010mq}, is
reproduced in the top panel of \reffig{data}.  The horizontal axis is
given in terms of the electron-equivalent energy $\Eee$ in units of
keVee, which is related to the nuclear recoil energy $\Enr$ by a
multiplicative quenching factor as discussed in \refapp{Quench}.  The
modulation amplitude cannot be given in terms of the nuclear recoil
energy in a model-independent way because the experiment does not
distinguish between sodium and iodine recoils on an event-by-event
basis and the recoil energy corresponding to a given
electron-equivalent energy, related by the nucleus-dependent quenching
factor, differs between the two nuclei.  In \reffig{data}, the
presence of a modulation is apparent below $\sim$6~keVee, while the
data above $\sim$6~keVee are consistent with zero modulation
amplitude.

Two possible WIMP masses can reasonably reproduce the observed
modulation amplitude spectrum: $\mchi\sim 10$~GeV (where sodium
recoils dominate) and $\mchi\sim 80$~GeV (where iodine recoils
dominate) \cite{Bottino:2003iu,Bottino:2003cz,
  Gondolo:2005hh,Petriello:2008jj,Chang:2008xa,Savage:2008er}.  The
predicted modulation spectrum for the two best-fit masses and
cross-sections, assuming the SHM and SI-only scattering, are shown in
\reffig{data}.  The behavior of the amplitude below the current
2~keVee threshold differs for the two WIMP masses, with the amplitude
of the heavy candidate going negative.  This is the phase reversal
feature in the annual modulation discussed in \refsec{SHM}.  Future
iterations of the DAMA experiment, which are expected to have lower
thresholds, should be able to distinguish these two possibilities.

The results of the DAMA experiment are in apparent contradiction with
the null results from other experiments, as will be discussed below.
Other conventional explanations for DAMA's observed annual modulation
have also been proposed~\cite{Schnee:2011ef}, including radon
contamination and neutrons~\cite{Ralston:2010bd}.  The modulating muon
flux has been studied as a potential contaminant in the
experiment~\cite{Nygren:2011xu, Schnee:2011ef, Blum:2011jf,
  Chang:2011eb, FernandezMartinez:2012wd}.  Thus far, most of these
explanations have been discounted (see Ref.~\cite{Bernabei:2012wp} for
a refutation of muons as a significant contaminant), but uncertainty
remains.

\subsubsection{\label{sec:CoGeNT}{The CoGeNT Experiment}}

The CoGeNT experiment, located in the Soudan mine in Minnesota,
consists of 440~grams of PPC Germanium detectors with a 0.4~keVee
energy threshold that makes it particularly well-suited to look for
light dark matter~\cite{Aalseth:2012if}.  Based upon 56~days of
exposure, the collaboration reported an excess of low energy events
above the well-known cosmogenic backgrounds~\cite{Aalseth:2010vx},
which could be consistent with a $\sim$10~GeV
WIMP~\cite{Fitzpatrick:2010em, Chang:2010yk}.  After more than a year
of data-taking, an annual modulation was reported at 2.8$\sigma$ with
a best-fit phase of April~16 \cite{Aalseth:2011wp} (see
Ref.~\cite{Arina:2011zh} for a Bayesian analysis).  The lower panel of
\reffig{data} shows the modulation amplitude observed in the CoGeNT
experiment for several energy bins, assuming the best-fit phase;
energies have been converted from electron-equivalent to nuclear
recoil enegies as described in \refapp{Quench}.  A significant
modulation is present above 5~keV, which is incompatible with the
total rate measured below 4~keV for standard assumptions about the
halo and scattering properties~\cite{Fox:2011px}.  However, the
modulation could be explained by local substructure~\cite{Fox:2011px,
  Kelso:2011gd, Natarajan:2011gz}.

\subsubsection{\label{sec:CDMS}{The CDMS Experiment}}

The CDMS experiment also consists of germanium and is located in the
Soudan mine. Using the ratio of two signals observed in an interaction
with the detector target---phonons and ionization---CDMS can
distinguish between nuclear recoil events (WIMP and/or neutron
interactions) and electron recoil events (beta and gamma
interactions), where the latter represents an otherwise dominant
background contribution~\cite{Akerib:2005zy}.  The conventional
low-background analyses in CDMS, the most recent having 612~kg-days of
exposure~\cite{Ahmed:2009zw}, have failed to detect any excess events
inconsistent with background.

To improve sensitivity to light WIMPs, which produce only low energy
recoils, CDMS has also performed a low-energy analysis
\cite{Ahmed:2010wy}, reducing its threshold from 10~keV to 2~keV.  At
these lower energies, it is more difficult to discriminate between
potential signal events and background events, so there is far more
background contamination in this analysis than the conventional case.
The $\sim$500 low-energy events found in this analysis are consistent
with rough background estimates.  However, CDMS places conservative
no-background-subtraction constraints on light WIMPs, neglecting
background contributions and allowing any or all of the events to be
due to WIMPs.  CDMS has also performed a modulation search in their
low-energy data \cite{Ahmed:2012vq}, finding no evidence for
modulation.  Constraints on the modulation amplitude, assuming the
CoGeNT best-fit phase of April~16, are shown in the lower panel of
\reffig{data}.  A direct comparison can be made with the CoGeNT
modulation results because both experiments have a germanium target,
although the CDMS modulation analysis was only performed down to
5~keV, whereas the CoGeNT modulation data goes to a much lower
$\sim$2~keV.

\subsubsection{\label{sec:CRESST}{The CRESST Experiment}}

The CRESST experiment, developed at the Max Planck Institute in Munich
and deployed in the Gran Sasso Tunnel, has 730~kg-days of data with a
CaWO$_4$ scintillating crystal target, and measures both light and
heat to reject electron recoils.  It reports an excess of low energy
events with a statistical significance of over 4$\sigma$
\cite{Angloher:2011uu}.  The experiment is not background-free
however, and has experienced problems with energetic alpha and lead
ions produced in the decay of polonium, itself produced from radon
decay.  For various technical reasons not discussed here, polonium
deposited in the clamps holding the detectors in place is the major
source of such backgrounds.  The expected number of these events,
which occur on the surface, is determined by extrapolating from high
energy observations (where such background events are readily
identifiable) to the signal regions at lower energies using Monte
Carlo simulations.  Questions have been raised as to whether the Monte
Carlo simulations underestimate the background contamination by
failing to account for the roughness of the surface at microscopic
scales~\cite{Kuzniak:2012zm}.  An upcoming redesign should eliminate
this background source, and future CRESST runs should clarify the
origin of the current excess.

\subsubsection{\label{sec:XENON}{The XENON Experiment}}

The XENON collaboration has developed a series of liquid xenon target
experiments, with the most recent iteration (XENON100) containing
$\sim$100~kg of xenon \cite{Aprile:2011dd}.  As with CDMS and CRESST,
XENON uses two signals---scintillation and ionization in this
case---to discriminate between nuclear recoils and electron recoils.
XENON100 and XENON10 both performed conventional
low-background analyses~\cite{Aprile:2012nq,Angle:2007uj}.  In addition,
XENON10 published a low-energy analysis that sacrifices background
discrimination to improve sensitivity to light WIMP
masses~\cite{Angle:2011th}.  None of these analyses find an excess of
events above expected background and XENON100 currently places the most
stringent constraints on the SI cross-section for WIMPs heavier than
$\sim$10~GeV.

\subsubsection{\label{sec:Other}{Other Experiments}}

The experiments we discussed in detail here represent only a fraction
of the current direct detection program.  No other experiment claims
an excess of events consistent with dark matter and, for standard
assumptions, none provide constraints as stringent as those from CDMS
and XENON100.  One exception is the case of SD scattering where the
coupling to the neutron is suppressed relative to the proton
($\anSD\ll\apSD$).  The proton-even target materials in CDMS and XENON
only couple weakly to the WIMP in this case (see \refeqn{sigmaSD}), so
these experiments place relatively weak constraints.  For this case,
COUPP \cite{Behnke:2012ys}, PICASSO \cite{Archambault:2012pm}, and
SIMPLE \cite{Felizardo:2011uw} provide the best limits.

\subsection{\label{sec:Compatibility} Compatibility of Experimental Results}

\begin{figure}
   \insertfig{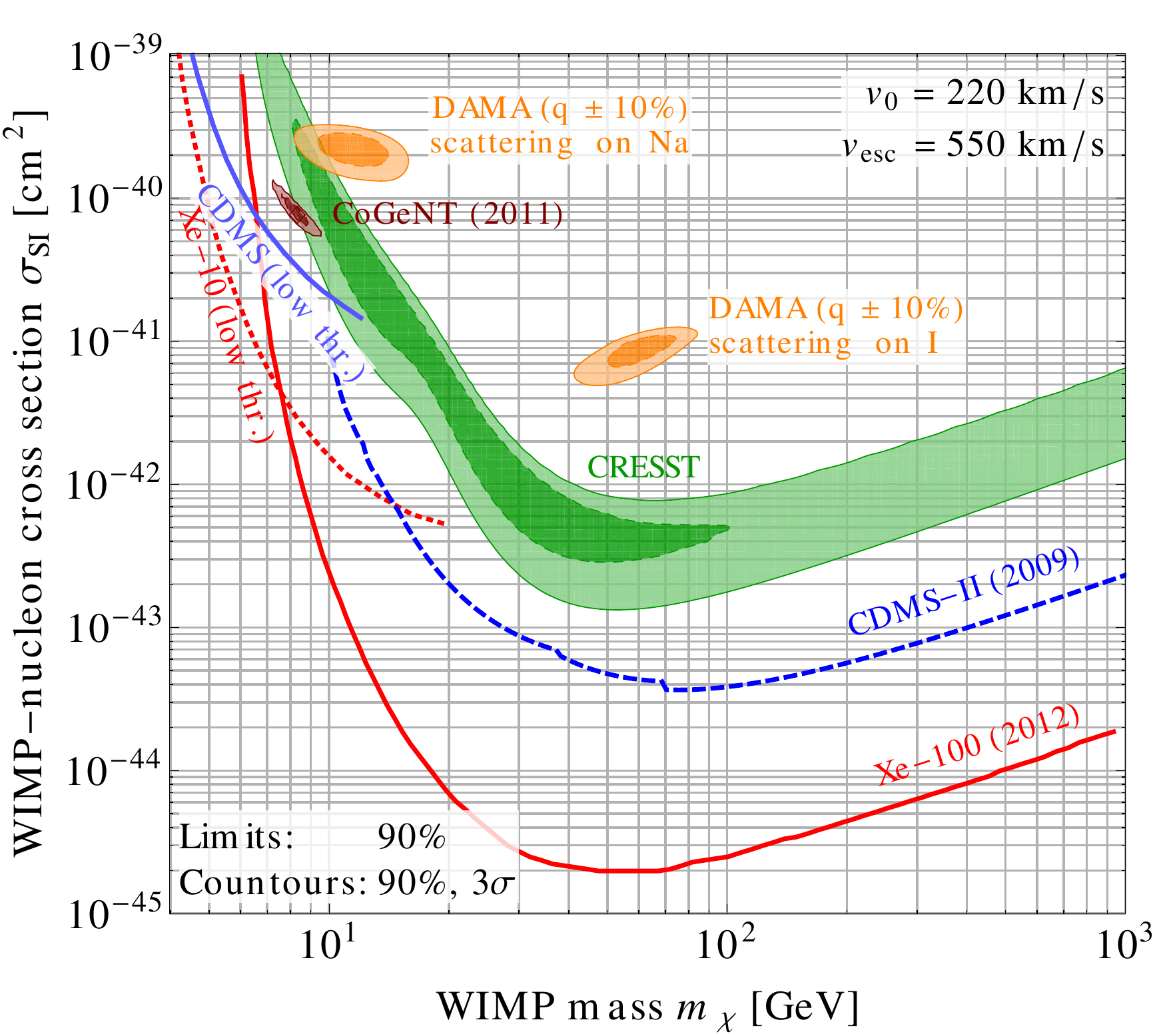}
  \caption{
    WIMP mass and SI cross-sections consistent with the anomalies seen
    by DAMA, CoGeNT, and CRESST, as well as constraints placed by the
    null results of CDMS and XENON (as of summer 2012).
    The halo model is assumed to be the SHM with the given parameters.
    The lack of overlap between the regions of the three anomalous
    results and their locations above the exclusion curves of CDMS and
    XENON indicate a conflict between the experimental results in this
    case.  Alternative couplings, modified halo models, and systematic
    issues have been proposed to reconcile this apparent incompatibility.
    Figure courtesy of J.~Kopp~\cite{Kopp:2011yr}.
    }
  \label{fig:msigma}
\end{figure}

\Reffig{msigma} summarizes the current status of anomalies and limits
for SI scattering, assuming the SHM with $\vmp = 220$~km/s and
$\vesc=550$~km/s.  Note that the experimental limits and anomalies, as
shown in this figure, are highly dependent on the assumptions made
about the particle and astrophysics (see~\cite{HerreroGarcia:2012fu,
  HerreroGarcia:2011aa,Fox:2010bz,Fox:2010bu} for
astrophysics-independent comparisons).  The compatibility may change
for \eg\ a different WIMP-nucleus effective operator or for additional
substructure contributions~\cite{Fairbairn:2008gz,Farina:2011pw,
Schwetz:2011xm,Frandsen:2011ts,Fornengo:2011sz, Arina:2011zh}.
However, note that changes to the particle physics and/or astrophysics
may change the interpretation of individual results without actually
affecting the compatibility among different results.\footnote{%
  Newer measurements of the Sun's velocity relative to the Galactic
  halo (as high as 250~km/s \cite{Reid:2009nj}, as opposed to the
  canonical 220~km/s in common use) shifts the best fit regions and
  the limit curves to the left.  For the SHM, the regions compatible
  with DAMA, CoGeNT, and CRESST move down in mass by a few
  GeV~\cite{Savage:2009mk}.  Because the bounds from the null
  experiments move to lower masses as well, the discrepancy between
  experiments is not alleviated.
  }
In addition, various systematic issues regarding the behavior of
individual detectors, such as the calibration of the recoil energy
scale, can impact the interpretation of experimental
results.\footnote{%
  Considerable discussion remains as to the true sensitivity of the
  XENON experiment near the energy threshold (see
  \eg~\cite{Collar:2010gg,Savage:2010tg}).
  }

The high-mass $\mathcal{O}$(80~GeV) DAMA region appears to be ruled
out for both SI and SD elastic scattering by null results from CDMS,
XENON, and COUPP---a heavy mass WIMP is only viable for non-standard
interactions.  On the other hand, the compatibility of light
$\mathcal{O}$(10~GeV) dark matter remains the subject of some debate.
For the case of SI scattering, these positive results are in apparent
contradiction with each other and with CDMS and XENON. Some authors
have nevertheless argued that some of the results could potentially be
reconciled (see \eg\ Refs.~\cite{Hooper:2010uy,Collar:2012ed}).

For the case of SD scattering, the DAMA lower mass region has until
recently remained compatible with all experiments \cite{Ullio:2000bv,
Savage:2004fn,Savage:2008er}, provided the SD coupling to the neutron
is strongly suppressed relative to the proton ($|\anSD| \ll |\apSD|$).
Results from PICASSO have since closed this window for standard
assumptions \cite{Archambault:2012pm}.
This particular case is also uniquely suited to be probed by indirect
searches involving detection of neutrinos produced by WIMPs annihilating
in the Sun, \eg\ with the Super-Kamiokande \cite{Desai:2004pq} and
IceCube detectors \cite{IceCube:2011aj}.

\subsection{\label{sec:Future} Future Prospects}

Direct detection experiments are poised at an important juncture.  In
the past few years, the cross-sections reached by the detectors have
improved by roughly two orders of magnitude.  A similar improvement is
expected in the next generation of detectors, which will be one tonne
(1000~kg) in size.  These experiments will probe some of the most
promising regions of WIMP parameter space, exploring Higgs exchange
cross-sections and large regions of supersymmetric parameter space.
However, as the sensitivity of direct detection experiments reaches
$\sigmapSI \sim 10^{-47}$~cm$^2$, astrophysical neutrinos become an
irreducible background, so the experiments are no longer
zero-background~\cite{Monroe:2007xp, Strigari:2009bq, Gutlein:2010tq}.
In addition to tonne-size detectors pushing the reach to lower cross
sections and heavier dark matter masses, efforts are also being made
to explore dark matter with masses below $\sim$1~GeV using electron
recoils~\cite{Essig:2011nj, Essig:2012yx}.

New technology and creative experimental designs will allow for
further exploration of the $\mathcal{O}(10 \text{ GeV})$ dark matter
anomalies.  For example, KIMS~\cite{Kim:2012rz} and
ANAIS~\cite{Amare:2011zz}, which use CsI(Tl) and NaI(Tl) targets,
respectively, will test the DAMA modulation claim.
\mbox{DM-Ice}~\cite{Cherwinka:2011ij} is a detector located at the South
Pole that also uses the same target material as DAMA.
Because it is located in the southern hemisphere and is embedded deep
in the ice where the natural temperature variation is minimal, DM-Ice
should have different environmental background sources than DAMA.

In addition, directional detectors will provide a powerful probe in
mapping out the distribution of the local dark matter.
Whereas the modulation in the recoil rate discussed throughout this
paper resulted from the variation in the \textit{velocity} of the
detector relative to the dark matter halo (due to the Earth orbiting the
Sun and, to a much lesser extent, the rotation of the Earth), detectors
with recoil direction sensitivity will observe a diurnal modulation in
the recoil direction due to the \textit{rotation} of the detector as
the Earth spins (\ie\ the orientation of the detector with respect to
the halo changes throughout the day).
The incoming WIMP flux is peaked in the direction of the Sun's motion
and, as a result, the nuclear recoil angular spectrum is peaked in the
opposite direction for most energies.  Therefore, the event rate should
experience a strong forward-backward (`head-tail') asymmetry along the
direction of the disk rotation.  In addition, the direction of the dark
matter wind as observed in the lab frame changes with the time of day
due to the Earth's daily rotation.  The result is a differential recoil
rate at a particular angle (as measured in the lab frame) that diurnally
modulates with an amplitude as large as $\sim$100\%~\cite{Spergel:1987kx,
Gondolo:2002np}, far larger than the modulation effects that are the
focus of this paper.
Ref.~\cite{Ahlen:2009ev} reviews the current status of prototypes of
directional detection experiments.  To achieve reasonable angular
resolution, the recoiling nucleus must leave a track that is
sufficiently long.  As a result, the chosen detector material is a
gas, typically CF$_4$ and CS$_2$ in current designs.  The use of gas
as the active target, with the gas being at low pressure (well below
atmospheric pressure to allow for longer recoil tracks), will require
these detectors to have volumes of $\orderof{10^4~\mathrm{m}^3}$ to
achieve tonne-scale masses.

A novel type of directional detector has also recently been proposed
that uses a DNA tracking material \cite{Drukier:2012hj}.  These
detectors can achieve nanometer resolution with an energy threshold of
0.5 keV and can operate at room temperature.  When a WIMP from the
Galactic halo elastically scatters off of a nucleus in the detector,
the recoiling nucleus then traverses thousands of strings of single
stranded DNA (ssDNA) and severs those ssDNA strings it hits.  The
location of the break can be identified by amplifying and identifying
the segments of cut ssDNA using techniques well known to biologists.
Thus, the path of the recoiling nucleus can be tracked to nanometer
accuracy.  By leveraging advances in molecular biology, the goal is to
achieve about 1,000-fold better spatial resolution than in
conventional WIMP detectors at a reasonable cost.

Directional detectors are particularly useful in mapping out the local
dark matter distribution~\cite{Copi:1999pw,Copi:2000tv,Alenazi:2007sy,
Bozorgnia:2011vc,Morgan:2004ys,Alves:2012ay,Lee:2012pf}.  A
positive signal at both a direct and directional detection experiment
would provide complementary information about the halo, building our
understanding of the velocity structure of the local dark matter.

\section{\label{sec:Summary} Summary}

The theoretical and experimental status of the annual modulation of a
dark matter signal (due to Earth's rotation around the Sun) in direct
detection experiments has been reviewed here.  Annual modulation
provides an important method of discriminating a signal from most
backgrounds, which do not experience such a yearly variation.  The
Milky Way halo consists of a dominant smooth component as well as
substructures such as streams, tidal debris, and/or a dark disk, each
of which contributes to the modulation of the signal.  In the Standard
Halo Model, the count rate in experiments should peak in June with a
minimum in December; substructure may change the phase, shape, and
amplitude of the modulation.  The current experimental situation is
puzzling, as several experiments have positive signals (DAMA and
CoGeNT both see annual modulation, while CoGeNT and CRESST-II have
unexplained events) but appear to be contradicted by null results from
other experiments (CDMS and XENON).  In the future, detectors with
sensitivity to the directionality of WIMPs should enable determination
of the direction of the WIMP wind as well as diurnal modulation due to
Earth's rotation.  Proposed techniques for directional detection
include large gaseous detectors as well as nanometer tracking with
DNA.  Consistent measurement of a head/tail asymmetry together with
annual modulation would provide very convincing evidence of WIMP
detection.  The future of dark matter searches is promising and an
annual modulation signal should play an important role in the
interpretation and confirmation of a potential WIMP signal.


\begin{acknowledgments}
  K.F. thanks M.~Valluri and M.~Zemp for useful conversations.
  K.F.\ acknowledges the support of the DOE and the Michigan Center for
  Theoretical Physics via the University of Michigan.
  K.F.\ thanks the Caltech Physics Dept for hospitality and support
  during her sabbatical.
  K.F.\ is supported as a Simons Foundation Fellow in Theoretical
  Physics.
  M.L.\ is supported by the Simons Postdoctoral Fellows Program and the
  U.S.\ National Science Foundation, grant NSF-PHY-0705682, the LHC
  Theory Initiative.   
  C.S.\ is grateful for financial support from the Swedish Research
  Council (VR) through the Oskar Klein Centre.
  C.S.\ thanks the Department of Physics \& Astronomy at the University
  of Utah for support.
  K.F.\ and M.L.\ acknowledge the hospitality of the Aspen Center for
  Physics, which is supported by the National Science Foundation Grant
  No.~PHY-1066293.
\end{acknowledgments}



\appendix

\section{\label{sec:Quench} Quenching Factor}

Any experimental apparatus does not directly measure the recoil energy
of scattering events.  The recoiling nucleus (or recoiling electron,
in the case of some backgrounds) will transfer its energy to either
electrons, which may be observed as \eg\ ionization or scintillation
in the detector, or to other nuclei, producing phonons and heat; these
are the signatures that are measured.  Some experiments that measure
only scintillation or ionization give their results in terms of the
electron-equivalent energy $\Eee$ of an event in their detector
(usually given in units of keVee).  This quantity is defined as the
energy of an electron recoil that would produce the observed amount of
scintillation or ionization, even if the event was actually a nuclear
recoil rather than an electron recoil.  Nuclear recoils tend to
produce a smaller amount of scintillation/ionization than electron
recoils for the same recoil energy, so $\Eee$ is \textit{not} the
recoil energy of that event if it is a nuclear recoil event.  For
nuclear recoils, these two energies are related by $\Eee = Q \Enr$,
where $Q$ is called the quenching factor.  The quenching factor is
different for each element in a detector and can have a recoil-energy
dependence.  The different quenching factors for different elements
and for electron-recoil events ($Q=1$ for electron recoils, by
definition) makes it impossible to determine the recoil energy of an
event based upon the scintillation or ionization signal alone.

Take, for example, the NaI in DAMA, with $Q_{Na} \approx 0.3$ and
$Q_{I} \approx 0.09$ \cite{Bernabei:1996vj} (see
Ref.~\cite{Tretyak:2009sr} and references therein for quenching factor
measurements of NaI and several other scintillators used in direct
detection experiments).  A recoil event in DAMA that produces 2~keVee
of scintillation can be from a $\sim$7~keV Na recoil, a $\sim$22~keV I
recoil, or a 2~keV electron recoil.  DAMA is unable to distinguish
between these three types of events on an event-by-event basis, so any
DAMA analysis is necessarily based upon the total $\dRdEee$ spectrum
that contains contributions from all three types of events.

CoGeNT, which observes only ionization in a germanium target, also
gives results in terms of the electron-equivalent energy spectrum.
CoGeNT suggests using $Q(\Enr) = 0.19935 \Enr^{\,0.1204}$ as a
reasonable approximation to the quenching factor measurements over the
energy range of interest \cite{Collar:2011pc} (measurements of $Q$ in
germanium can be found in the Appendix of Ref.~\cite{Lin:2007ka}).
CDMS, which also uses a germanium target, can discriminate between
electron-recoil and nuclear-recoil events and can reconstruct the
nuclear recoil (though limited by a finite resolution), so results for
this experiment are given in terms of $E \approx \Enr$ rather than
$\Eee$.  Because CDMS and CoGeNT are made of the same target material,
one would expect that these two experiments should have the same
nuclear recoil spectrum; however, one should keep in mind the caveat
that the two results are given in terms of different quantities that
must be rescaled to make direct comparisons between their two results.

\section{\label{sec:MIV} Mean Inverse Speeds of Commonly Used Velocity
                Distributions}

The detection rate in dark matter experiments is directly proportional
to the mean inverse speed $\eta(\vmin) = \int_{|\bv|>\vmin} d^3v \,
\frac{f(\bv,t)}{v}$.  Here, we present analytical results of this
integration quantity for several commonly used isotropic velocity
distributions.  We define $\tilde{f}(\bv)$ as the velocity
distribution in the rest frame of the dark matter population (\ie\
$\int d^3v \, \bv \tilde{f}(\bv) = \mathbf{0}$).  The velocity
distribution in the lab frame is determined via the Galilean
transformation $f(\bv) = \tilde{f}(\bvobs+\bv)$, where $\bvobs(t)$ is
the (time-dependent) motion of the lab (observer) relative to the rest
frame of the dark matter population.  This motion is described in more
detail in \refsec{Modulation}.

The distributions considered below are Maxwellian distributions
(including two modifications to account for a finite cutoff) and
distributions corresponding to cold flows and debris flows.  These are
not the only possible distributions of dark matter and may be only
simple approximations for some populations, but they are frequently
used distributions that have known analytical forms for $\eta$.

\subsection{\label{sec:Maxwellian} Maxwellian Distributions}

Perhaps the most useful simple distribution is the Maxwellian:
\begin{equation} \label{eqn:Maxwellian}
  \widetilde{f}(\bv) =
      \left( \frac{1}{\pi \vmp^2} \right)^{3/2} \, e^{-\bv^2\!/\vmp^2} \, .
\end{equation}
For this distribution,
\begin{equation} \label{eqn:etaMaxwellian}
  \eta(\vmin,t) = \dfrac{1}{2 \vobs} \Big[ \erf(x+y) - \erf(x-y) \Big] \, ,
\end{equation}
where $\vmp$ is the most probable speed,
\begin{equation} \label{eqn:xy}
  x \equiv \vmin/\vmp \quad \textrm{and} \quad
  y \equiv \vobs/\vmp \, .
\end{equation}
Many well-mixed populations of dark matter particles can be expected
to have a Maxwellian or Maxwellian-like distribution, in which case
the above is a useful first approximation.

The Standard Halo Model (SHM) takes the dark matter halo to be an
isothermal sphere, in which case the velocity distribution is
Maxwellian.  However, high velocity particles would escape the Galaxy,
so the high-velocity tail of the distribution is cut off in a
realistic halo model.  \refsec{SmoothHalo} presents two methods for
removing the tail of the Maxwellian, with the resulting distributions
given by Eqns.~(\ref{eqn:TruncMaxwellian})
\&~(\ref{eqn:SubMaxwellian}).  For the SHM or any dark matter
component described by one of these two velocity distributions, the
mean inverse speed $\eta$ is \cite{Savage:2006qr,McCabe:2010zh}
\begin{equation} \label{eqn:etaMaxwellianCutoff}
  \eta(\vmin) =
    \begin{cases}
      \quad\, \dfrac{1}{\vobs}
        \qquad\qquad\qquad\qquad\qquad\qquad\qquad\qquad\qquad
        & \textrm{for} \,\, z<y, \, x<|y\!-\!z| \, , \\[2ex]
      \multicolumn{2}{l}{\!
      \dfrac{1}{2 \Nesc \vobs}
        \left[
          \erf(x\!+\!y) - \erf(x\!-\!y)
           - \frac{4}{\sqrt{\pi}} \left(1-\beta(x^2\!+\!\tfrac{1}{3}y^2\!-\!z^2)\right) y e^{-z^2}
        \right] } \\
        & \textrm{for} \,\, z>y, \, x<|y\!-\!z| \, , \\[1ex]
      \multicolumn{2}{l}{\!
      \dfrac{1}{2 \Nesc \vobs}
        \left[
          \erf(z) - \erf(x\!-\!y)
          - \frac{2}{\sqrt{\pi}} \left(y\!+\!z\!-\!x -\tfrac{1}{3}\beta(y\!-\!2z\!-\!x)(y\!+\!z\!-\!x)^2 \right) e^{-z^2}
        \right] } \\
        & \textrm{for} \,\, |y\!-\!z|<x<y\!+\!z \, , \\[1ex]
      \quad\;\;\; 0
        & \textrm{for} \,\, y\!+\!z<x
    \end{cases}
\end{equation}
where $\beta=0$ for \refeqn{TruncMaxwellian} and $\beta=1$ for
\refeqn{SubMaxwellian}, $x$ and $y$ are as defined above, and
\begin{equation} \label{eqn:z}
  z \equiv \vesc/\vmp \, .
\end{equation}
Note that the normalization factor $\Nesc$ has a different form for
the two distributions.

\subsection{\label{sec:ColdFlow} Cold Flow}

Cold flows, such as tidal streams, have small to negligible velocity
dispersions.  In the case of zero dispersion,
\begin{equation} \label{eqn:ColdFlow}
  \tilde{f}(\bv) = \delta^3(\bv)
\end{equation}
and
\begin{equation} \label{eqn:etaColdFlow}
  \eta(\vmin) = \dfrac{1}{\vobs}\, \theta(\vobs - \vmin) \, ,
\end{equation}
where $\theta$ is the Heaviside function.  Although the velocity
dispersion is small in cold flows, in some cases such as tidal
streams, it is not completely negligible.  In those cases, a useful
approximation can often be made with the Maxwellian distribution of
\refeqn{Maxwellian} with a small $\vmp$.

\subsection{\label{sec:DebrisFlow} Debris Flow}

To first order, the debris flow in a Milky Way-like galaxy
\cite{Lisanti:2011as} has an isotropic, constant-speed velocity
distribution in the Galactic rest frame that can be described by a
delta function in \textit{speed}, as opposed to the delta function in
\textit{velocity} seen with cold flows above:
\begin{equation} \label{eqn:DebrisFlow}
  \tilde{f}(\bv) = \frac{1}{4\pi\vflow^2} \delta(|\bv|-\vflow) \, ,
\end{equation}
where $\vflow$ is the uniform speed of the particles.  In the lab
frame \cite{Kuhlen:2012fz},
\begin{equation} \label{eqn:etaDebrisFlow}
  \eta(\vmin) =
    \begin{cases}
      \dfrac{1}{\vflow} \, ,
        & \textrm{for} \,\,  \vmin < \vflow\!-\!\vobs \\[2ex]
      \dfrac{1}{2\vflow\vobs} \left[ \vflow + \vobs - \vmin \right] \, ,
        & \textrm{for} \,\,  \vflow\!-\!\vobs < \vmin < \vflow\!+\!\vobs \\[2ex]
      0 \, ,
        & \textrm{for} \,\,  \vflow\!+\!\vobs < \vmin  \, .
    \end{cases}
\end{equation}
Although the debris flow has both a dispersion in the speed and some
anisotropy in the Galactic rest frame, the above reference shows that
this analytical form still provides a reasonable approximation.


\end{document}